\documentclass[preprint,12pt]{elsarticle}

\usepackage{amssymb}
\usepackage{amsmath}

\usepackage{color}
\newcommand{\ansA}[1]{{\color{black}#1}}

\journal{Journal of Computational Physics}

\begin{document}

\begin{frontmatter}



\title{Advection of the image point in probabilistically-reconstructed phase spaces}


\author{Igor Shevchenko} 

\affiliation{organization={National Oceanography Centre},
            addressline={European Way},
            city={Southampton},
            postcode={SO14 3ZH},
            country={UK}}

\begin{abstract}
Insufficient reference data is ubiquitous in data-driven computational fluid dynamics, as it is usually too expensive to compute or impossible to observe over long enough times needed for data-driven methods.
The lack of data can significantly compromise the fidelity of results computed with data-driven methods or render them inapplicable.
To challenge this problem, we propose a probabilistic reconstruction  method that enhances the
hyper-parameterisation (HP) approach with ideas underlying the probabilistic-evolutionary approach.
We offer to use the HP method ``Advection of the image point'' on data sampled from the joint probability distribution of the reference dataset.

The HP method has been tested regionally on the sea surface temperature and surface relative vorticity computed with the global 1/4$^\circ$ and 1/12$^\circ$ resolution NEMO model.
Our results show that the HP solution (the solution computed with the HP method) in the probabilistically-reconstructed and reduced (in terms of dimensionality) phase space at 1/4$^\circ$ resolution is more accurate
than the 1/4$^\circ$-solution computed with NEMO.
Additionally, the HP solution is several orders of magnitude faster to compute than the 1/4$^\circ$ NEMO solution.
The proposed method shows encouraging results for the NEMO model and the potential for the use in other operational
ocean and ocean-atmospheric models for both deterministic and probabilistic predictions.
Furthermore, the method can be used as a fast reanalysis tool allowing the complex dynamics of a comprehensive ocean model to be approximated by the HP solution.
It can also function as a dynamic interpolation method to fill gaps in observational data.
\end{abstract}



\end{frontmatter}

\section*{Plain Language Summary}
The shortage and damage of reference data used for training or calibrating data-driven methods
is a common issue in ocean modelling. This problem can greatly undermine the accuracy of the
methods or make them unusable.
We propose a solution to mitigate this problem and a hyper-parameterisation (HP) method that
can utilize short and damaged data from comprehensive ocean models
otherwise falling outside its range of applicability.
The methodology proposed in this study
not only enables HP to work with sparse and damaged data sets but also serves as a fast reanalysis tool;
it approximates complex ocean flows with HP solutions, which are several orders of magnitude
faster to compute. Additionally, the proposed methodology
can be used as a dynamic interpolation technique to fill gaps in observational data
for further use in reanalysis and data assimilation.

%
%

%


%
%
%
%

\section{Introduction}
The data-driven paradigm is gradually gaining momentum in computational fluid dynamics by offering
an alternative to the physics-driven approach in situations when there is no physics-based model for
the studied phenomenon or it is too computationally expensive to acquire data from the model
(e.g.~\cite{mendez_ianiro_noack_brunton_2023} and references therein).
The data-driven approach includes a variety of methods, spreading
across different fields (e.g.
statistical modelling~\cite{StorchZwiers2002,VanemZhuBabanin2022},
dynamical system reconstruction~\cite{AguirreLetellier_2009,Brunton_et_al_2016,Brunton_et_al_2017,MangiarottiHuc_2019},
hybrid modelling~\cite{SB2022_J2,SC2024_J1},
probabilistic-evolutionary approach~\cite{SB2023_J2}, etc.)
In recent years, machine learning (ML) methods started to enrich the data-driven approach
with artificial neural networks (e.g.~\cite{Schneider_et_al2017,ZhangLin2018,GormanDwyer2018,ZB2020,Dramsch2020,BNK_2022})
typically trained on reanalysis data or
reference runs of physics-driven models.
The majority of these ML methods are used
for predictive modelling (e.g.~\cite{Ham_et_al2019,Bi_et_al2023,Kochkov_et_al_2024}), while some, known as Interpretable ML,
go beyond traditional ML predictions and try to improve our understanding
of modelled phenomena (e.g.~\cite{Adadi_Berrada2018,Betancourt_et_al2022}).

Being a powerful tool providing
a wide spectrum of methods ranging from purely-data to physics-data driven, also known as hybrids,
the data-driven approach typically suffers from the lack of reference data that
can significantly compromise its fidelity or make it inapplicable. In order to help this problem,
we offer a probabilistic reconstruction method which augments the hyper-parameterisation (HP) approach
(e.g.,~\cite{SB2021_J1,SB2022_J2,SB2022_J1,SB2023_J1,SC2024_J1}) with some ideas underlying the probabilistic-evolutionary approach~\cite{SB2023_J2}.
More specifically, we propose to use Advection of the image point (a data-driven HP method working in phase space, also
called state space) together with
data sampled from the joint probability distribution of reference data
so as to probabilistically reconstruct the reference phase space for the HP method and make it operable
for short or damaged (unreliable for further use)
reference data records otherwise falling out of its range of applicability.
\ansA{
The main advancements of this study relative to the previous HP works are:
the introduction of a probabilistic phase-space reconstruction layer based on JPD sampling; a dimensionally efficient strategy that samples only state distributions while computing tendencies locally; demonstration of robustness to severely damaged reference datasets; application in a high-dimensional ocean modelling context via reduced EOF representation.
}
The main difference between the proposed methodology and ML methods (based on artificial neural
networks) is that: the former does not have the training phase (the calibration procedure of the HP
method might be
interpreted as such though), it uses phase spaces instead of the physical space,
respects the regional stability property,
and relies on the unconventional calculation of directional vectors.
The regional stability is defined as a property of the solution of a differential equation
to permanently remain in a neighborhood of the phase space region if the initial condition is
sufficiently close to that region. By the unconventional calculation of directional vectors we mean
how the right-hand side (RHS) in the HP method is calculated. Namely, we do not use a sequence
of following every other states from the past (as conventional integrators) to calculate the RHS, but
use the states neighboring the current state in the phase space.

Although we consider the use of probabilistically-reconstructed phase spaces within the HP framework,
its application is not bound to HP alone. For example, probabilistically-reconstructed
reference datasets can be used within the context of other data-driven methods,
including the training of artificial neural networks and reconstructing
missing parts in observational data from different sources (drifters,
weather stations, satellites, etc.)

The term \textit{probabilistic}, used throughout this study, refers to two distinct but complementary
aspects of the method.
First, the HP framework is probabilistic in the sense that it is grounded in the
empirical estimation of the joint probability distribution (JPD).
Local averages used in the method, and optionally second moments,
are computed from this distribution using neighborhood-based statistics.
While the resulting dynamics is deterministic once the reference phase space is fixed,
its structure is governed entirely by the statistical geometry of the data.
The evolution is deterministic, but it is driven by empirical estimates derived from a probability
distribution, not from first-principles physics.
Second, the method is probabilistic in the stricter sense that, when the reference phase space is
sparse or contains voids, it can be reconstructed or augmented by sampling from the JPD.
This probabilistic filling is performed offline, prior to trajectory integration, and is crucial in
enabling the HP method to operate in regions poorly represented by the raw reference data.
However, once the reference phase space is built the HP trajectory evolves deterministically.
In this way, the probabilistic nature of the method lies in both
the statistical geometry of the phase space and the probabilistic reconstruction of under-sampled regions,
but not in the runtime dynamics itself.

\ansA{it 's also important to note that the method does not explicitly identify voids in phase space. Instead, it relies on sampling from the estimated JPD, which increases density in statistically under-sampled regions. Void filling is therefore implicit and governed by the estimated probability density. Regions with zero empirical support cannot be reconstructed.
}

\section{The probabilistic reconstruction method: schematic and basics}
\begin{figure}[h]
\hspace*{-1.25cm}
\includegraphics[scale=0.275]{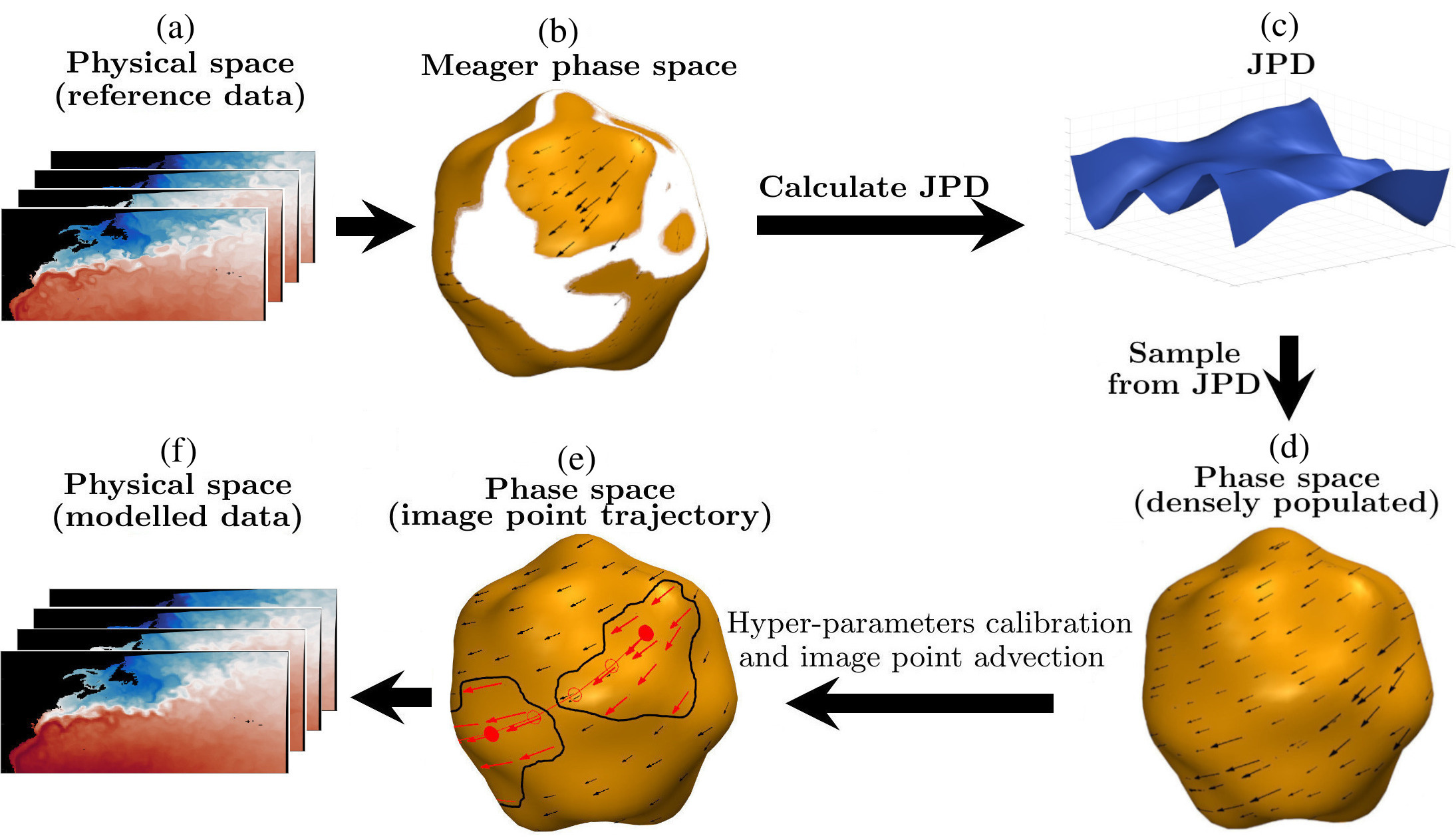}
\caption{Schematic of the probabilistic reconstruction method.}
\label{fig:schematic}
\end{figure}

The idea underlying the probabilistic reconstruction method is to calculate the
joint probability distribution (JPD) for a given reference dataset and then
sample reference data from this JPD instead of using the reference dataset itself.
Hence, the JPD becomes an extra layer between the reference data and a data-driven method
thus providing an extra source of reference data from the reference data distribution.
It would be instructive to first look into the proposed method's schematic
(figure~\ref{fig:schematic}) to better understand its specifics and then consider
the method in more detail.

{\bf (a) Reference data acquisition}. The first step is to provide the method with reference data. Depending on the problem to solve, it can be numerical simulations, observations, or both. In our case it is sea surface temperature (SST) computed with a global 1/12$^\circ$ resolution
ocean model NEMO and then interpolated onto a 1/4$^\circ$ grid (figure~\ref{fig:schematic}a);
SST is a scalar field defined on a cartesian grid.
If observations are available then they can also be included in the reference dataset by simply adding them to the
reference data record. In order to be consistent in space, observations must be on the same grid as the NEMO data.
\ansA{
It is important to note that two distinct roles of time must be clearly separated in the methodology:

{\bf 1. Preprocessing stage (derivative estimation).} The reference tendencies $F(x_i)$ are estimated using a central finite-difference approximation, which requires temporally ordered data. Temporal consistency is therefore essential during this preprocessing step.

{\bf 2. HP trajectory reconstruction stage.} Once state-tendency pairs $(x_i,F(x_i))$ are constructed, the HP method operates purely in phase space. The reference data are treated as an unordered cloud of points, and chronological ordering is no longer used during trajectory integration.
}

{\bf (b) Reference phase space calculation}.
Before proceeding with the second step it would be helpful to remind the reader that
a phase space (or state space) is an abstract mathematical space of all possible states of a physical system, while the physical space is typically defined as the actual three-dimensional space we all live in.
\ansA{In this study, a discretized scalar field defined on a spatial grid with $n$
grid points is represented as a vector in $\mathbb{R}^n$, so that each grid point corresponds to one coordinate of the phase space. This grid-point representation is not essential to the method. More generally, the HP framework operates in an abstract state space and allows for arbitrary coordinate transformations of the data. For example, one may represent the system via EOF-PC decomposition, wavelet transforms, spectral expansions, or other linear or nonlinear mappings. The grid-point formulation used here is therefore a special case of a broader class of admissible phase space representations.}

The second step is to get the reference phase space for the HP method called
"Advection of the image point"~\cite{SB2021_J1,SB2023_J1};
the reference phase space is shown as an orange blob in figure~\ref{fig:schematic}b.
The latter needs states (which is SST itself) and directional vectors
(time derivatives of SST, also referred to as tendencies), which are calculated
from the reference data; different schemes can be used, we use the central finite difference in time.
The tendencies are shown as black vectors in figure~\ref{fig:schematic}b, and the points they are attached to represent SST,
which is a vector in the phase space, while it is a scalar field in the physical space.

The white spots (which we call voids) are regions of the phase space
which lack data (usually, due to short reference records or damaged observations).
Typically, these voids are a reason why data-driven methods cease to work on meager phase spaces.

{\bf (c) Joint Probability Distribution calculation}. The third step is to calculate the joint probability distribution (JPD) from the reference data in phase space;
it is shown as a blue surface in figure~\ref{fig:schematic}c for illustration purposes.
In high-dimensional phase spaces, such as the one used for SST, the JPD is a hyper-surface.
It is important to remark that we calculate the JPD solely for SST, while SST tendencies are
computed in a different way explained below. The JPD for the reference SST is calculated globally,
meaning that the entire reference dataset is used in this calculation.

The standard approach to calculate an approximation to the JPD is to use the histogram method.
This method is based on the subdivision of the phase space
into hypercubes (bins) and counting the number
of states in each bin to compute the height of each column of the multidimensional histogram.
The method works well only in low-dimensional spaces, since the computer memory
needed grows as the number of bins to the power of the space dimensions.
Therefore, in high-dimensional
spaces (like the one used for SST) computing a multidimensional histogram
and keeping it in memory for further sampling is an unaffordable option.

In this study, we do not calculate the whole multi-dimensional histogram.
Instead, we compute a marginal probability density function (PDF), or coordinate-wise PDF,
for each component of the state (i.e. for each grid point of SST)
and then use these PDFs to sample from the JPD;
it gives us access to all necessary information for sampling at virtually no cost.
Notice that these PDFs can be interpolated if
there is insufficient reference data to accurately approximate them.

\begin{figure}[h]
\centering
\includegraphics[scale=0.3125]{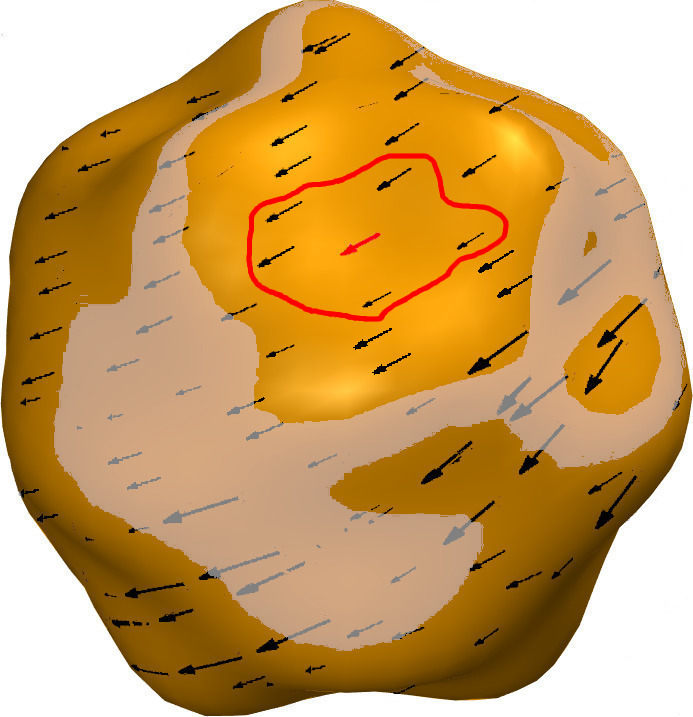}
\caption{Shown is a schematic representation of the probabilistically-reconstructed phase space with the superimposed
voids (grey regions). A new tendency (red vector) is computed as an average of reference
tendencies (black vectors) in the neighborhood of sampled SST (red contour).
}
\label{fig:jpd_rhs}
\end{figure}

{\bf (d) Sampling from JPD and calculation of probabilistically-reconstructed phase space}.
In the fourth step, we sample SST from the JPD and then calculate SST tendencies by averaging
reference SST tendencies over the neighborhood of sampled SST (figure~\ref{fig:jpd_rhs}).

This neighborhood-localized calculation of SST tendencies allows to decrease
the dimensionality of the JPD two times (as the dimension of JPD for both SST and its tendencies is two times larger
compared to the JPD only for SST) and thus makes its calculation more efficient without compromising on accuracy.
Put differently, sampling SST fields and their associated tendencies jointly would require
estimating the full JPD in a space of dimension  $2n$, where $n$ is the number of spatial grid points
used to discretize the SST field. In contrast, we sample only from the JPD of SST fields
(which is of dimension $n$), and then calculate SST tendencies locally.
This reduces the required dimensionality of the JPD two times (from $2n$ to $n$),
since the tendencies are not independently sampled but are functionally linked to the sampled SST
states via local averaging.

It is worth mentioning that tendencies can also be computed in a probabilistic
way from a local JPD which is constructed based on the information from the neighborhood of the new tendency
(red contour in figure~\ref{fig:jpd_rhs}), but this method is slower than the one used in this study.

In order to sample from the JPD we use the inverse transform sampling method~\cite{Devroye1986}.
We also tried the rejection sampling method but did not gain that much of a difference in its favour;
the sampling method based on a reverse-time stochastic differential equation~\cite{Sungbin_et_al2023}
might be another option instead of the inverse transform sampling.
Sampling from the JPD allows us to probabilistically reconstruct the originally sparsely-populated reference phase space by filling it with new states and directional vectors (tendencies) needed to advect the image point
(figure~\ref{fig:schematic}d). Note that the probabilistically-reconstructed reference data can be used not only in the HP method, but
in other data-driven methods which suffer a lack of data.

We would like to draw the reader's attention to the fact that the new states generated by sampling from the JPD
does not mean new reference states (i.e. the states generated by the underlying
reference model, it is the NEMO model in our case), since
sampling from the JPD cannot guarantee that sampled states are reference states, and
the HP dynamics in probabilistically-reconstructed phase spaces cannot guarantee it either.

\ansA{
To explain how the sampling procedure calculates a new state, we consider a
three-dimensional case consistent with Chua's circuit -- a set of points in an $XYZ$-space.
We sample new points from the joint probability distribution $P(x,y,z)$ by decomposing it
into a marginal and conditional distributions:

\[
P(x,y,z) = P_X(x)\,P_{Y|X}(y \mid x)\,P_{Z|X,Y}(z \mid x,y).
\]

First, the marginal distribution $P_X(x)$ is estimated from the available data and converted
into a cumulative distribution function (CDF), $F_X(x)$. A uniform random number
$u \in [0,1]$ is then drawn, and the new $x$-coordinate is obtained by inverting the CDF:

\[
x_{\mathrm{new}} = F_X^{-1}(u).
\]

Next, the conditional distribution $P_{Y|X}(y \mid x_{\mathrm{new}})$ is determined, again
converted into a CDF $F_{Y|X}(y)$, and a second uniform random number
$u' \in [0,1]$ is drawn to produce:

\[
y_{\mathrm{new}} = F_{Y|X}^{-1}(u').
\]

Finally, the conditional distribution $P_{Z|X,Y}(z \mid x_{\mathrm{new}},y_{\mathrm{new}})$ is
estimated, converted into a CDF $F_{Z|X,Y}(z)$, and a third uniform random number
$u'' \in [0,1]$ is drawn to compute:

\[
z_{\mathrm{new}} = F_{Z|X,Y}^{-1}(u'').
\]

This sequential inverse transform sampling procedure is repeated to generate as many new
points $(x_{\mathrm{new}},y_{\mathrm{new}},z_{\mathrm{new}})$ as required, ensuring that the samples
follow the original 3D JPD without explicitly constructing a full 3D JPD, thereby avoiding
the higher memory cost associated with direct estimation in three dimensions.

%
}

It is also important to note that SST and its tendencies in the probabilistically-reconstructed phase space are not ordered in time.
This is the moment when we need Advection of the image point to compute a trajectory (solution), i.e to order those points in time.

{\bf (e) Advection of the image point in probabilistically-reconstructed phase space}. In the fifth step, the probabilistically-reconstructed phase space is used by Advection of the image point to calculate the trajectory (red line in figure~\ref{fig:schematic}e).
For the reader's convenience, we briefly describe how Advection of the image point works. This method falls into the category of data-driven methods
from the hyper-parametersation (HP) class which takes advantage of working in phase spaces, as opposed to the conventional methods operating in the physical space. The method has been tested on both idealized and comprehensive ocean models (two-layer quasi-geostrophic model in a channel and MITgcm in the North Atlantic configuration) and showed significant improvements of the HP solution toward the reference one~\cite{SB2021_J1,SB2023_J1}.
The HP approach currently umbrellas five different methods (ranging from purely data-driven to hybrids, which combine the data- and physics-driven paradigms).
Another striking feature of the HP approach is that its measure of goodness is how close the HP solution is to the reference phase space.
To put it another way, the HP approach matches phase spaces (the reference phase space and the phase space where the HP solution evolves), whereas conventional methods match individual trajectories.
This measure allows the HP solution to evolve in the neighborhood of the reference phase space, and therefore reproduce the flow dynamics
which is very similar to the reference one.

The idea behind Advection of the image point (which we refer to as the HP method in what follows) is to use reference data, say $\mathbf{x}\in\mathbb{R}^n$, and the following differential equation

\begin{equation}
\frac{d\mathbf{y}}{dt}=\frac{1}{M}\sum\limits_{j\in\mathcal{U}_J}\mathbf{F}(\mathbf{x}_j)+
\eta\left(\frac{1}{N}\sum\limits_{i\in\mathcal{U}_I}\mathbf{x}_i-\mathbf{y}(t)\right),\quad
\mathbf{y}(t_0)=\mathbf{y}_0,
\label{eq:evolution_eq_nudging}
\end{equation}
to describe the evolution of the image point, $\mathbf{y}\in\mathbb{R}^n$,
in the phase space of reference data. In the context of this work, $\mathbf{x}(t)$ is the probabilistically-reconstructed SST,
and $\mathbf{y}(t)$ is the HP solution computed in the probabilistically-reconstructed phase space,
and $\eta$ is the nudging strength. 
The neighbourhood of
$\mathbf{y}(t)$ is denoted as $\mathcal{U}(\mathbf{y}(t))$,
$\mathcal{U}_I$ and $\mathcal{U}_J$ are the sets of timesteps indexing the discrete reference solution $\mathbf{x}$ in $\mathcal{U}(\mathbf{y}(t))$. Although the method allows to use different sets of reference tendencies $\mathbf{F}(\mathbf{x}_j)$
and points $\mathbf{x}_i$, we set $M=N$ and $\mathcal{U}_I=\mathcal{U}_J$ in this study.

In a nutshell, the actual dynamics is given by the observed reference tendency
$\displaystyle\frac{1}{M}\sum\nolimits_{j\in\mathcal{U}_J}\mathbf{F}(\mathbf{x}_j)$
the imperfect reconstruction of which from the reference data is compensated by nudging towards the observed reference state
$\displaystyle\frac{1}{N}\sum\nolimits_{i\in\mathcal{U}_I}\mathbf{x}_i$
in the probabilistically-reconstructed phase space. Note that
the neighbourhood is computed as the average over
$M$ (and $N$ for the nudging term) nearest, in $l_2$ norm, to the solution $\mathbf{y}(t)$ points.
The choice of the norm and the way the neighborhood $\mathcal{U}(\mathbf{y}(t))$ is calculated is not limited to those used in this study, and can vary depending on what is needed. Our choice is, probably, the simplest, but it serves well the purpose of this work.
We dub the set of parameters $\{M,N,\eta\}$ hyper-parameters.
The hyper-parameters can be set based on the chosen metric and available data, we will return to their choice later.
The interested reader is referred to~\cite{SB2021_J1,SB2023_J1} for a more detailed discussion of the method.

\begin{figure}[h]
\centering
\includegraphics[scale=0.3]{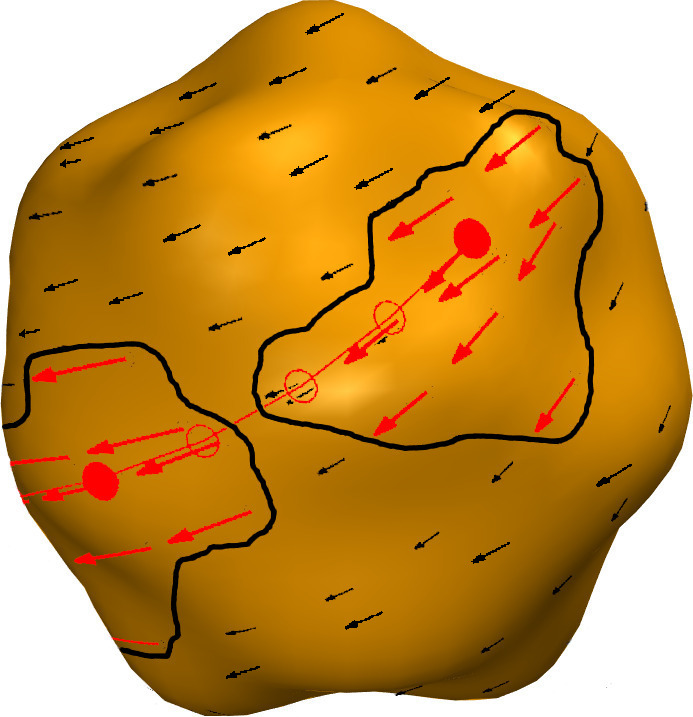}
\caption{Shown is a sketch of the probabilistically-reconstructed reference phase space (orange blob),
vectors $\mathbf{F}(\mathbf{x}_i)$ (black arrows) pointing out from points $\mathbf{x}_i$.
For the HP method all these points is a cloud of points which are not ordered in time.
The HP solution $\mathbf{y}(t)$ is shown as a red trajectory.
\ansA{The filled red circle denotes the current state $\mathbf{y}(t)$ at time $t$, while the empty red circles represent other points along the HP trajectory. The neighborhood $\mathcal{U}(\mathbf{y}(t))$ is illustrated by the black contour surrounding the filled red circle.
}
The red vectors (and also each point they are attached to) within the neighborhood are
the $N$ reference fields $\mathbf{x}_i,\, i\in\mathcal{U}_I$ (and $M$ reference tendencies $\mathbf{F}(\mathbf{x}_i)$) nearest, in $l_2$-norm, to the HP solution $\mathbf{y}(t)$. We assume
that $\mathcal{U}_I=\mathcal{U}_J$ and $M=N$, but the method is not bound to this particular choice.}
\label{fig:ref_space}
\end{figure}

The characteristic feature of the HP method is that it looks at the reference dataset
as being a cloud of points which are not ordered in time in contrast to traditional methods, which
rely on the chronological order. It allows the HP method to constantly stay in the reference
phase space, as there always exists a neighborhood of nearest points.
Thus, the HP method never runs out of reference data.
This is why we write $\mathbf{x}_i$ and $\mathbf{F}(\mathbf{x}_j)$ instead of $\mathbf{x}(t)$ and
$\mathbf{F}(\mathbf{x}(t))$ in equation~\eqref{eq:evolution_eq_nudging}.
At each time step, the HP method takes $N$ nearest $\mathbf{x}_i$ (and $M$ tendencies $\mathbf{F}(\mathbf{x}_j)$)
to the solution $\mathbf{y}(t)$ points (figure~\ref{fig:ref_space}).
The main advantage of using the HP method in probabilistically-reconstructed phase spaces is that the data for the method
is sampled from the same distribution as the reference data itself.

It is important to differentiate between the reference data set and
the reference phase space. The former is the skeleton of the latter, i.e. the reference data set is a set of states used in the proposed
methodology to discover the reference phase space through the use of the HP method in the probabilistically-reconstructed phase space.

Another important aspect of the HP method is that
only first moments of the local reference
neighborhood are used: the mean position $\overline{\mathbf{x}}$
and the mean tendency $\overline{\mathbf{F}}(\mathbf{x})$.
This yields a trajectory driven by the average local flow and attracted towards the centroid of
nearby states, thus offering a stable, low-variance approximation. However, this formulation is
agnostic to the geometry of the neighborhood, as it treats all directions equally and fails to exploit
any local anisotropy or alignment in the data. Including second-moment information,
such as the covariance matrix of positions $\mathbf{C}_\mathbf{x}$
and the covariance of tendencies $\mathbf{C}_\mathbf{F}$
introduces important geometric structure. For example, the nudging term can be replaced
with
$$
\eta\mathbf{C}^{-1}_{\mathbf{x}}\left(\frac{1}{N}\sum\limits_{i\in\mathcal{U}_I}\mathbf{x}_i-\mathbf{y}(t)\right)
$$
to modulate correction strength based on local variance, and the mean tendency can be refined
by projecting $\mathbf{F}$
along the principal directions of $\mathbf{C}_\mathbf{F}$,
thus leading to directionally aware transport, i.e.
rather than treating all directions equally, the method adapts to the local flow geometry
by steering the trajectory along the locally dominant dynamical pathway.
This enhances reconstruction fidelity in
phase-space regions where the reference neighborhood is locally stretched, sheared, or curved, which is
a common situation in GFD. The downside is that second moments are more
susceptible to sampling noise, can be ill-conditioned and require matrix inversion,
thus increasing computational cost and risking numerical instability.
Furthermore, if the neighborhood is undersampled or nearly isotropic, second-moment corrections may
amplify noise without improving accuracy. In short, while the first-moment HP equation is robust,
efficient, and broadly applicable, the second-moment extension provides anisotropic, geometry-aware
refinement, which is beneficial when the local structure is well-resolved and physically meaningful,
but risky when data is sparse or noisy.

{\bf Calibration of hyper-parameters}.
The HP method has hyper-parameters $\{M,N,\eta\}$ which should be properly set to address the problem at hand.
In order to calibrate these hyper-parameters we define a measure of discrepancy of two clouds of points, say $A$ and $B$, as follows

\begin{equation}
\mathrm{Di}(A,B):=\|A-\{\widehat{\min}\, d(a_i,B)\}_{1\le i\le |A|}\|_2/\|A\|_2\,,
\label{eq:discrep}
\end{equation}
where $d$ is the Euclidean distance between the $i$-th point in $A$ and a point in $B$;
$|\cdot|$ is the number of points in the cloud (it is assumed that $|A|=|B|$);
$\|\cdot\|_2$ represents the $l_2$-norm.
The cloud A represents the reference data, while the cloud B is the HP solution computed in the probabilistically-reconstructed phase space; the clouds are not supposed to have any order defined on them.

It is important to remark that the $\widehat{\min}$ in~\eqref{eq:discrep} is excluding, i.e. once the minimum between a pair $(a_i,b_j)$ has been found,
the point $b_j$ is excluded from $B$ (to avoid multiple counting) and is then used in comparison with element $a_i$.
In other words, $\widehat{\min}$ does not return the minimum distance between $a_i$ and $B$ per se, it returns the element $b_j$ that delivers
this minimum distance.
Also note that the choice of this measure of discrepancy reflects the measure of goodness
used for the HP method (namely, the proximity of the phase space
(where HP solution evolves) to the reference phase space, or more accurately to the
probabilistically-reconstructed reference phase space).
The calibration procedure consists of minimizing the discrepancy on the space of hyper-parameters;
the smaller the discrepancy is, the closer the clouds are to each other.
We would like to remind the reader again that the HP method tries to match phase spaces (not individual trajectories), and this is why the calibration of hyper-parameters makes itself useful.


{\bf (f) Evolution in physical space}.
The fifth step is to compute the modelled solution back in the physical space (figure~\ref{fig:schematic}f).
In this study, the solution computed with the HP method is reshaped from a vector (used in the phase space) into a scalar field, which is used to present fields in the physical~space.

\section{The probabilistic reconstruction method in action}
In this section we apply the proposed method to Chua's circuit to
demonstrate how it works in a low-dimensional phase space and then consider its application in the context of the global NEMO~model.

\subsection{Chua's circuit}
As a minimum working example, we consider Chua's circuit and demonstrate in more detail how the method works. Chua's circuit~\cite{CKM1986} is given
by the following system of equations:

\begin{equation}
\frac{d\mathbf{x}}{dt}=\mathbf{F}(\mathbf{x}(t)),\quad \mathbf{F}:=
\begin{pmatrix}
\alpha(y-ax^3-cx)\\
x-y+z\\
-\beta y - \gamma z\\
\end{pmatrix},
\label{eq:chua}
\end{equation}
with $\mathbf{x}(t)=(x(t),y(t),z(t))$, and $\alpha=10$, $\beta=15$, $\gamma=0.01$, $a=0.1$, $c=-0.2$.
As an initial condition, we take $\mathbf{x}(t_0)=(1,0,0)$ so that the solution is close to the attractor. 
Note that the solution $\mathbf{x}(t)$ at time $t$ (being voltages across the capacitors and the
current through the coil) is in the physical space, while the whole evolution of the solution (together with the time derivatives of the solution) is in the phase space.

{\bf Reference data acquisition}.
Within the context of Chua's circuit, the reference data for the HP method is the solution $\mathbf{x}(t)$ and its tendencies $\mathbf{F}(\mathbf{x}(t))$.
In order to get the reference data we integrate the Chua system~\eqref{eq:chua} over time $t\in[0,100]$ and save the solution.

\begin{figure}[h]
\vspace*{-0.5cm}
\hspace*{-1.75cm}
\includegraphics[scale=0.3125]{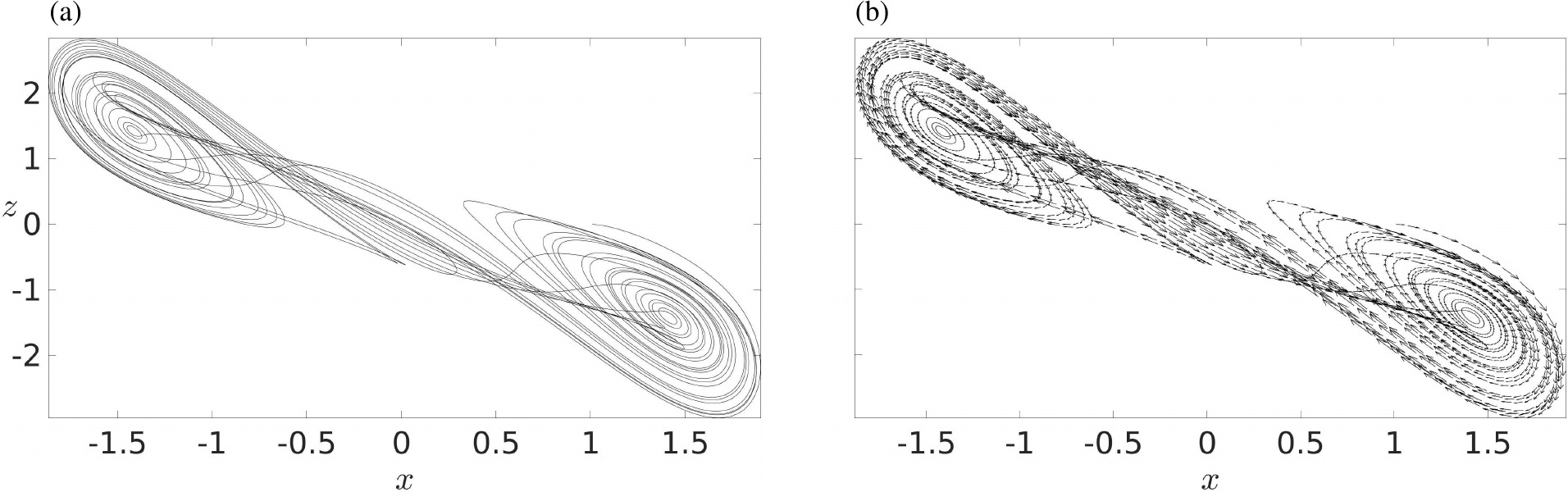}
\caption{Shown is (a) the solution of Chua's system $\mathbf{x}(t)$ and
(b) the corresponding vector field $\mathbf{F}(\mathbf{x}(t))$ for~$t=[0,100]$.}
\label{fig:chua}
\end{figure}

\noindent
Then, the tendencies are computed by differentiating $\mathbf{x}(t)$; we use the central finite difference in time (figure~\ref{fig:chua}).
\ansA{Although the right-hand side is explicitly known for Chua’s circuit, we do not use it in this study. Instead, we estimate tendencies via central finite differences applied to the numerical solution in order to  ensure methodological consistency with realistic applications, where the exact right-hand side may not be available. The HP method is designed to operate solely on state-tendency pairs extracted from data, without relying on knowledge of the underlying equations.}
Also note that instead of or in combination with the numerical solution, one can also use observations.
In this case, one should form a new dataset
$\widehat{\mathbf{x}}=\{\mathbf{x}(t),\mathbf{x}_o(t)\}$, which includes both the solution $\mathbf{x}(t)$ and
observations~$\mathbf{x}_o(t)$.
However, using observations in the HP approach is not just adding observed points into the phase space,
unless they lie on the same attractor or in its near vicinity (it is doubtful due to model biases,
resolution differences, or unresolved physics), they are noise-free (it is never true),
they respect the dynamical constraints of the simulation model (it is rare). Put differently,
simply injecting observed states into the phase space is insufficient and potentially destructive.

To prevent mismatches between simulations and observations, and
a possible drift of the HP solution towards the observational subspace,
an additional nudging term should be introduced in equation~\eqref{eq:evolution_eq_nudging}
to steer trajectories towards observations with a tunable strength;
this gives the control over how strongly one ''trusts'' the observations versus the numerical simulations.
When properly balanced against the existing nudging strength $\eta$, this hybrid formulation
yields trajectories that should lie between observed and simulated clouds, enabling dynamics
that respects both sources. The method's behaviour depends critically on the geometry, density, and
consistency of these two clouds, and on the choice of hyper-parameters.
A weighted-neighbourhood formulation is recommended to balance the influence.

Also, observed tendencies can be included in equation~\eqref{eq:evolution_eq_nudging}.
However, one should keep in mind that observed tendencies are often noisy,
sparse, or inconsistent with the model. The safest route is to use the tendencies from the numerical simulations,
while using observations to influence the trajectory via the observation-based nudging term.
But, if one does add observed tendencies they should be prperly weighted and smoothed to mitigate their
possibly negative effect.

\begin{figure}[h]
\hspace*{-1.5cm}
\includegraphics[scale=0.285]{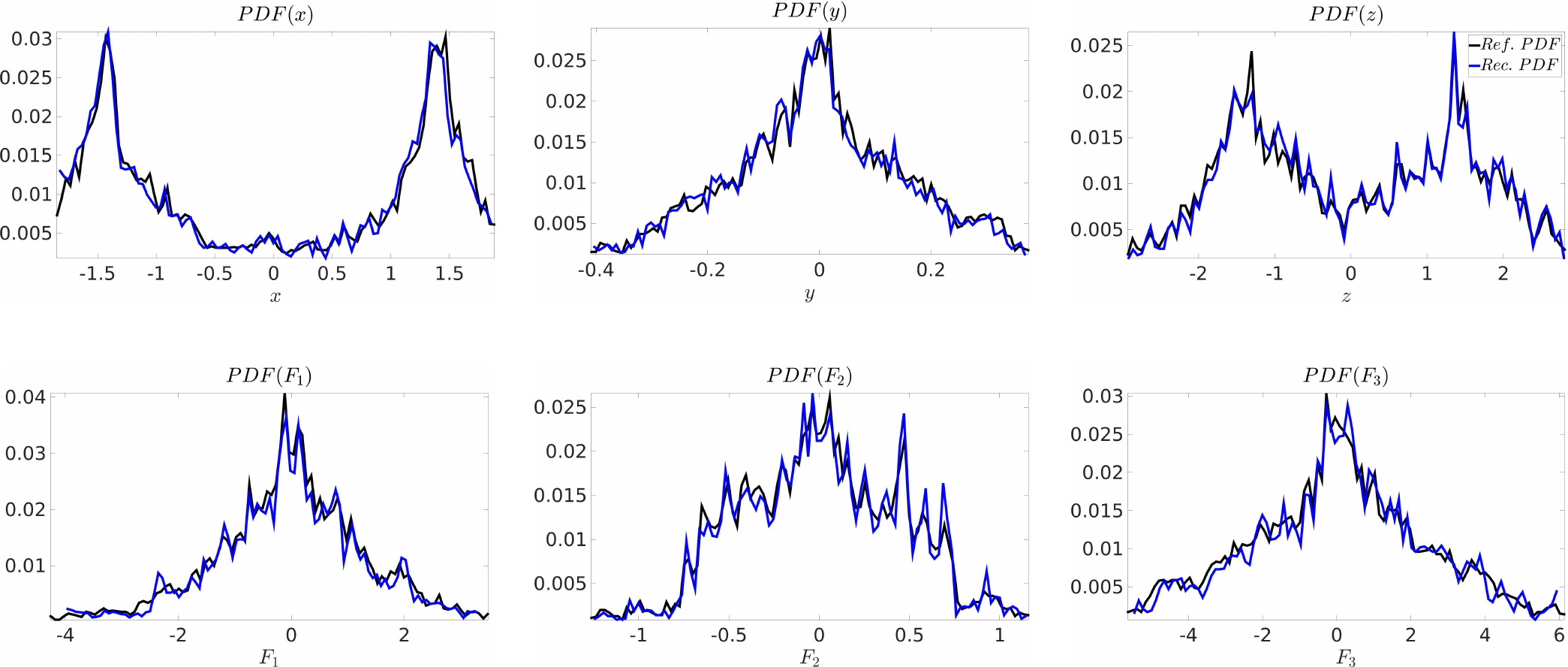}
\caption{Shown is the marginal reference PDF (black) and reconstructed PDF (blue)
for the solution of Chua's system (top row) and for the corresponding vector field (bottom row).
Note that the reconstructed PDF is an accurate approximation to the reference PDF.}
\label{fig:chua_ref_pdf}
\end{figure}

{\bf Calculation of JPD and sampling from JPD}.
The JPD of the reference solution $\mathbf{x}(t)$ and reference tendencies $\mathbf{F}(\mathbf{x}(t))$ are calculated as described above.
To ensure that the sampling from the JPD is sufficiently accurate for further use,
we compare the reference PDF (computed from the reference dataset)
and the reconstructed PDF (computed from the data sampled from the JPD).
As seen in figure~\ref{fig:chua_ref_pdf}, the reconstructed PDF (blue) is an accurate approximation to the reference PDF (black).
While it is possible to calculate the reconstructed PDF with higher accuracy,
doing so is not necessary for the purpose of this~study.

\begin{figure}[h]
\hspace*{-1.5cm}
\includegraphics[scale=0.23]{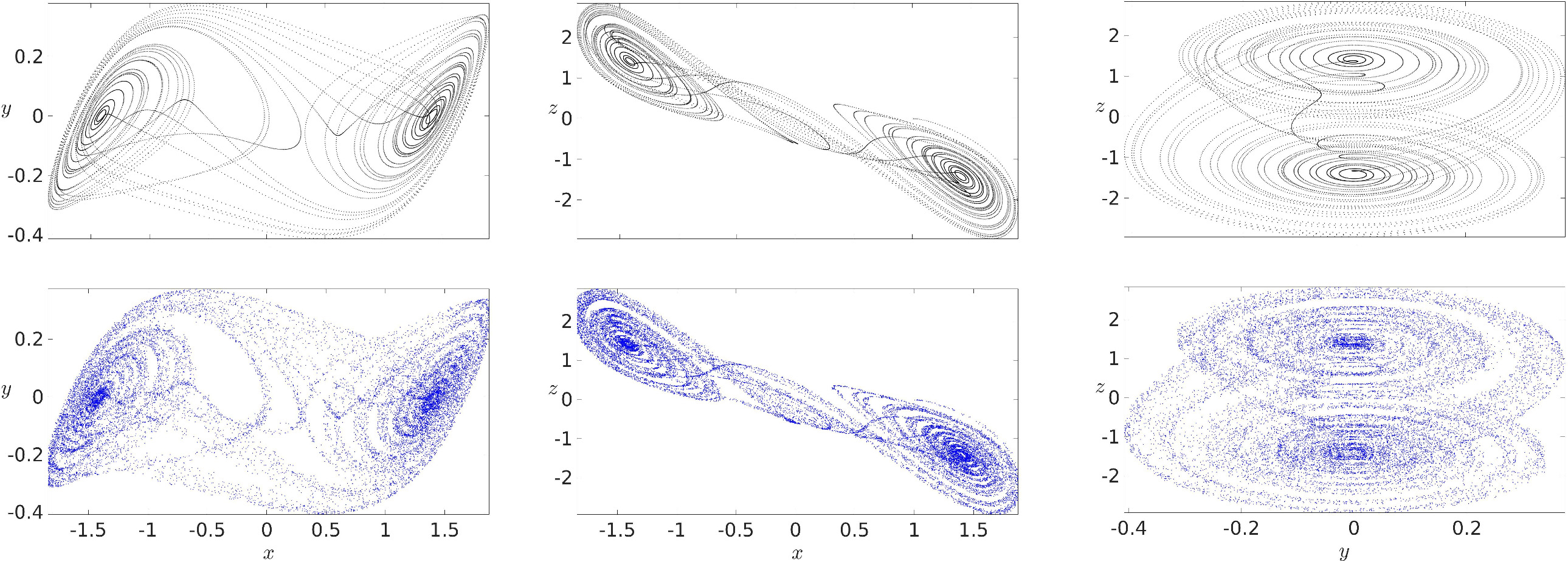}
\caption{Shown are different projections of the reference solution (top row) and its reconstruction from the JPD (bottom row).
The reconstructed data not only reproduces the reference one but, more importantly, densifies the phase space by adding new points,
as we sampled twice as much points from the JPD compared with the reference data.}
\label{fig:chua_ref_prob_sol}
\end{figure}

\begin{figure}[h!]
\hspace*{-1.5cm}
\includegraphics[scale=0.285]{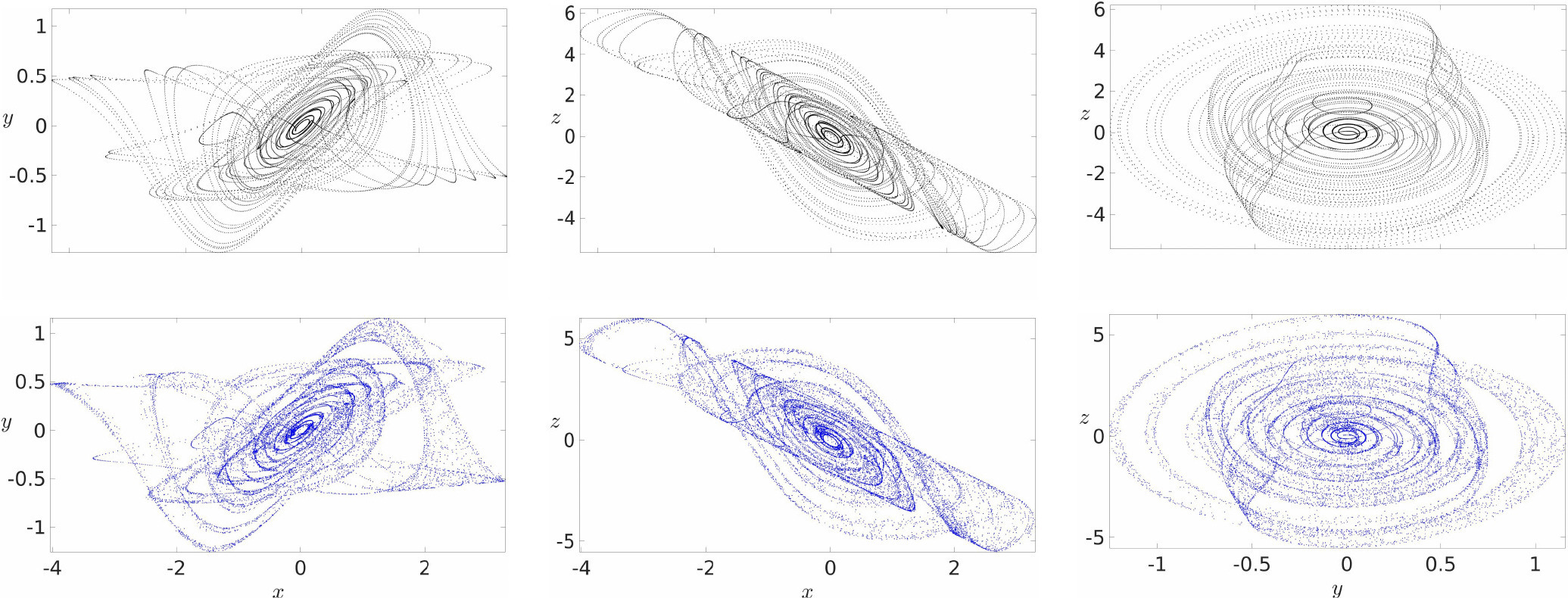}
\caption{The same as in figure~\ref{fig:chua_ref_prob_sol}, but for the tendencies. The tendencies
are not sampled directly from the JPD but are computed locally in the neighbourhood of the sampled solution.}
\label{fig:chua_ref_prob_rhs}
\end{figure}

The next step is to sample from the JPD to probabilistically reconstruct the reference phase space for the HP method.
We use the same amount of reference data as before, but harvest twice as much from the JPD (figures~\ref{fig:chua_ref_prob_sol},\ref{fig:chua_ref_prob_rhs}).
\noindent
The results clearly demonstrate not only the high quality of the probabilistically-reconstructed data, but also
that the reconstructed data covers a larger phase space compared with the reference one.
In other words, sampling from the JPD reconstructs the reference space geometry thus
implying correct dynamical behaviour of the HP method, i.e. the solution has the right range of
variability and local curvature, not just the right mean or energy.

\subsection{Calibration of hyper-parameters}
The reconstructed dynamics in phase space, computed with the HP method,
depends on the hyper-parameters $M$, $N$, $\eta$. Our studies (not shown) allow us to report the following
findings. Small values of $M$ and $N$ yield high-variance estimates of local means and tendencies,
which in turn can produce noisy, over-dispersed trajectories and inflated attractors.
Conversely, large $M$ and $N$ lead to over-smoothed reconstructions, suppressing local variability and
causing the attractor to shrink unnaturally. The nudging parameter $\eta$
controls the strength of the restoring term that pulls the trajectory toward the local mean:
small $\eta$ allows freer movement through phase space and results in broader, more variable attractors, while
large $\eta$ imposes strong corrections that constrain the dynamics, often collapsing the trajectory onto
the reference one. The balance of these parameters governs the trade-off between preserving
dynamical richness and maintaining numerical stability in the reconstruction.

In order to properly adjust these hyper-parameters, one can calibrate them by
minimizing $\mathrm{Di}(A,B)$ in the space of $M$, $N$, and $\eta$. One can opt for engaging a full-blown optimization to
find the optimum solution. However, for the purpose of this study, we set $M=N=5$
and \ansA{perform a one-dimensional grid search over $\eta\in[0,1]$, selecting the value that minimizes $\mathrm{Di}(A,B)$.}
We have found that $\eta=0.235$ gives a solution with the lowest discrepancy.
Once M and N are fixed, the solution for a given $\eta$ is not necessarily the one that ensures
the global minimum, only a local one.
Currently, only the accuracy in reconstructing the reference phase space is considered.
But in principle, one could also optimize for temporal smoothness of the HP trajectory,
physical consistency, forecast skill, robustness to perturbations, etc.

\begin{figure}[h]
\hspace*{-1.5cm}
\includegraphics[scale=0.285]{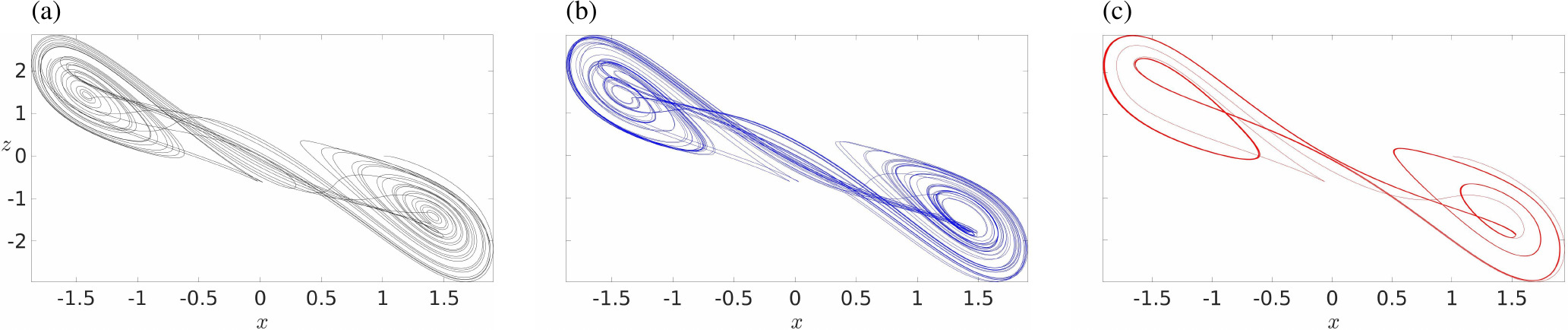}
\caption{Shown are (a) the reference solution, (b) HP solution computed in the probabilistically-reconstructed reference phase space,
and (c) HP solution computed in the reference phase space (no probabilistic reconstruction used); both HP solutions are computed for $M=N=5$ and $\eta=0.235$.
The HP solution computed in the probabilistically-reconstructed reference phase space is much more developed
compared with the HP solution computed in the reference phase space. Note that both HP solutions run for $t\in[0,200]$, i.e. 50\% of the time out of the reference
sample.}
\label{fig:chua_ref_prob_hp}
\end{figure}

In order to demonstrate that the proposed method can gain from using the
probabilistically-reconstructed phase space, we compare it with the HP method run on the reference dataset (figure~\ref{fig:chua_ref_prob_hp}).

As seen in figure~\ref{fig:chua_ref_prob_hp}b, the HP solution computed in the probabilistically-reconstructed phase space is very-well developed
on the attractor compared with the HP solution computed on the reference dataset (figure~\ref{fig:chua_ref_prob_hp}c), which remains bounded to
a very narrow band due to insufficient data.
All these results give an extra layer of confidence that the proposed method has strong potential for the use in
more sophisticated models which we consider in the next section.

\subsection{Damaged reference data}
\begin{figure}[h]
\hspace*{-1.5cm}
\includegraphics[scale=0.21]{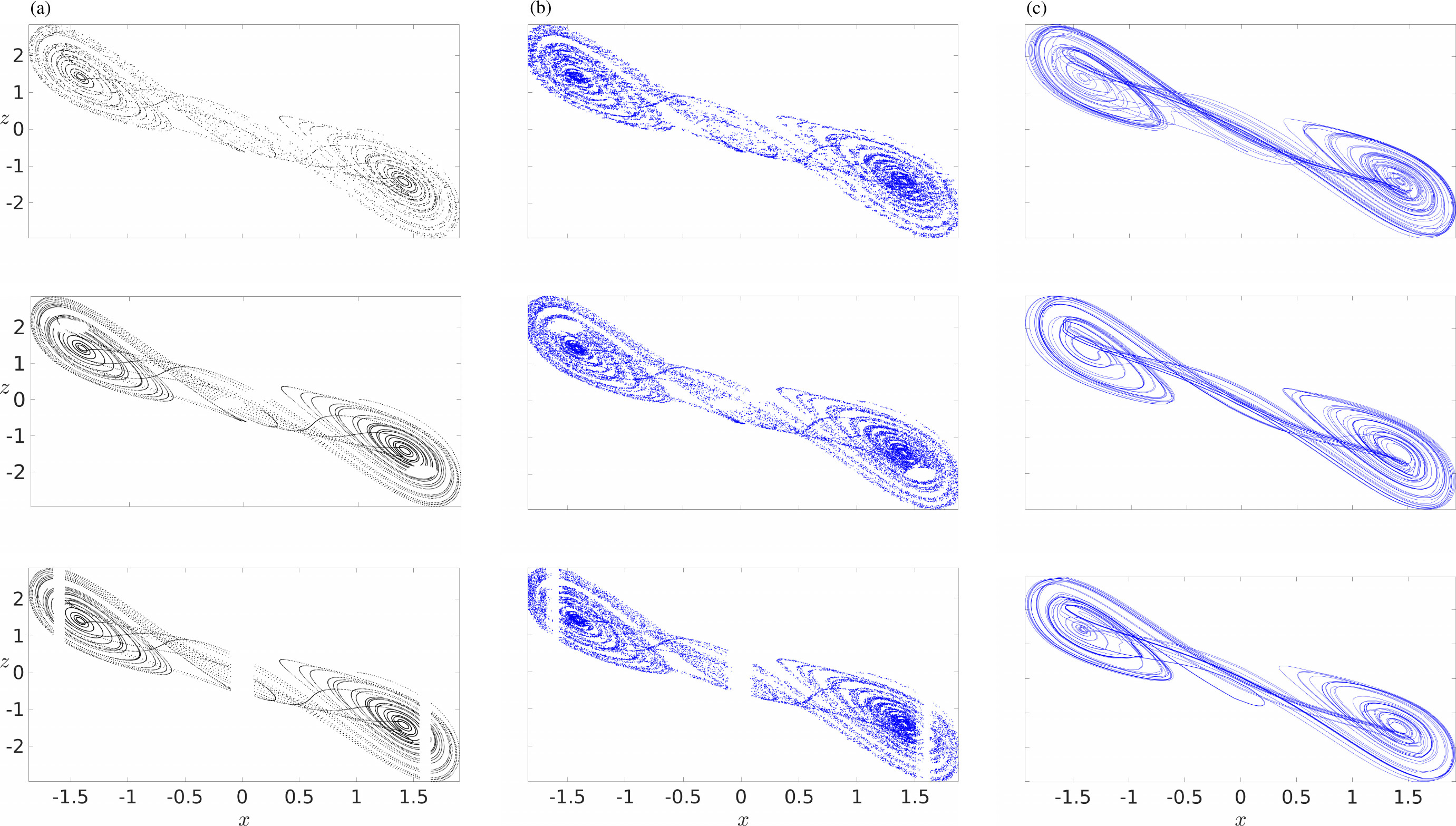}
\caption{Shown are (a) reference data sets with different level of damage (top -- (1) 50\% of randomly-chosen states has been removed from the
original reference data shown in figure~\ref{fig:chua_ref_prob_hp}a, middle -- (2) original reference data with gaps, bottom -- (3) original reference data with three cuts);
(b) phase spaces probabilistically-reconstructed from the corresponding damaged reference data sets; (c) HP solutions computed in the probabilistically-reconstructed
phase spaces. In all cases, the HP method can reproduce the solution on the attractor even when sampling from the JPD cannot
completely repair the damage. Note that for the HP method we use $\eta=0.003$ for cases (1) and (2), and $\eta=0.051$ for case (3).
}
\label{fig:chua_corrupted}
\end{figure}

Reference data with unreliable or missing states (which are called damaged in this study)
is ubiquitous in the realm of data-driven methods.
Reference data sets lacking states in the phase space or having vast voids (regions of no data)
can significantly influence the accuracy of data-driven methods or make them inapplicable.
Therefore, it is instructive to study what happens when
the reference data is damaged. As an example, we consider three different cases of
the reference phase space damaged with: (1) randomly-missed states, (2) gaps,
and (3) cuts (figure~\ref{fig:chua_corrupted}).
In all cases, sampling from the JPD reconstructs the phase space up to the extent,
which is enough for the HP method to work.
In the first case, sampling from the JPD densifies the attractor well (top subplot in figure~\ref{fig:chua_corrupted}b),
\ansA{because the global support of the attractor (the bounded region in phase space within which all trajectories of the dynamical system ultimately reside and evolve) remains intact, and only the sampling density is reduced.}
In the second and third cases, however, there are still some voids in the reconstructed phase space
(middle and bottom subplots in figure~\ref{fig:chua_corrupted}b) which cannot be completely filled in by sampling from the JPD, \ansA{because the global support of the
attractor is significantly distorted by removed coherent regions of the attractor},
but the HP method does reproduce the dynamics on the attractor (similar to the reference one) nonetheless.

The ability of the HP method to work in damaged phase spaces can be exploited
as a fast reanalysis method in which
the complex dynamics of a comprehensive ocean model is replaced with the HP solution.
Essentially, this method can be thought of as a dynamic interpolation, which can stitch gaps
in observational data using the HP solution.

\ansA{
On the other hand, linear or spline interpolation can fill gaps between known samples too,
but these methods operate purely in the static data space and do not account for the
underlying dynamics of the system. In contrast, the HP method reconstructs
missing states by integrating an explicit dynamical model in the reconstructed
phase space, guided by the joint probability structure of the reference data.
This ensures that the reconstructed trajectories:
\begin{enumerate}
    \item \textbf{Remain within the reference phase space}, thus avoiding excursions into dynamically
    forbidden regions that static interpolation can produce;
    \item \textbf{Respect the local flow geometry}, as the HP update uses neighbouring
    states in phase space to determine both the direction and speed of evolution;
    \item \textbf{Handle sparse or irregular sampling}, i.e. cases in which simple interpolation
    becomes ill-defined or distorted.
\end{enumerate}
In the low-dimensional example in figure~\ref{fig:chua_corrupted}, spline or linear interpolation could visually
connect points, but such reconstructions would not match the true time evolution
over longer intervals, particularly in cases with under-sampling or in
higher-dimensional settings such as SST or SRV, where the reference space geometry is much more
complex.

The HP method produces dynamically consistent trajectories because it does not
interpolate states purely in a geometric sense. Instead, the method integrates an
explicit evolution equation in phase space, whose right-hand side is constructed
from local dynamical information in the reference dataset. Specifically, the HP
update uses the tendencies $F(x_j)$ of neighbouring states $x_j$ in phase space,
together with a relaxation term towards nearby observed states. These neighbours
are selected so that they correspond to states actually visited by the system,
ensuring that the reconstructed fields are consistent with the reference flow dynamics.
By integrating this data-driven dynamical system, the HP
method advances the state along directions observed in the reference dynamics, thereby
avoiding excursions into dynamically forbidden regions. In contrast, spline or
linear interpolation merely connects points geometrically, without enforcing
consistency with the underlying vector field, and can therefore produce
trajectories that are not dynamically realisable.
}

Before we proceed to the application of the method to the NEMO model, it is instructive to clarify
a subtle point:
how the method, although demonstrated to work with data from an autonomous dynamical system,
can be equally applicable to data originating from a non-autonomous ocean model.
This is possible because the HP method is fundamentally time-agnostic, as it operates in phase space,
which contains no explicit time coordinate. Whether the system is autonomous or non-autonomous
is irrelevant to the method's mechanics. What matters is that the reference dataset contains a
representative sample of the system's possible states, including those generated under
time-varying forcings. If so, the HP trajectory can reconstruct dynamical behavior from
the geometry of the reference phase space. Time enters only indirectly, through
how it shapes the distribution of states, not through any explicit appearance in the formulation.
However, if the reference phase space does not contain sufficient coverage of the system's evolution
under time-varying forcings, HP cannot reconstruct this dynamics.

\subsection{NEMO model~\label{sec:res}}
The output of the global 1/4$^\circ$ (ORCA025-N4001) and 1/12$^\circ$ (ORCA0083-N006) resolution NEMO model
is used in this study.
The reference solution is the 5-day mean sea surface temperature (SST) and surface relative vorticity (SRV) in the North Atlantic region
$[-83^{\circ}W,-20^{\circ}W]\times[27^{\circ}N,55^{\circ}N]$ over the period of 1979-1993;
the period of 1994-2007 is used for validation to check how the HP method performs. The reference solution
is interpolated from the 1/12$^\circ$-grid onto the 1/4$^\circ$ grid.
The lower-resolution 1/4$^\circ$-solution from ORCA025-N4001 simulation (we call it
the modelled solution) is used for comparison with the HP solution.

\begin{figure}[h]
\hspace*{-2cm}
\includegraphics[scale=0.3]{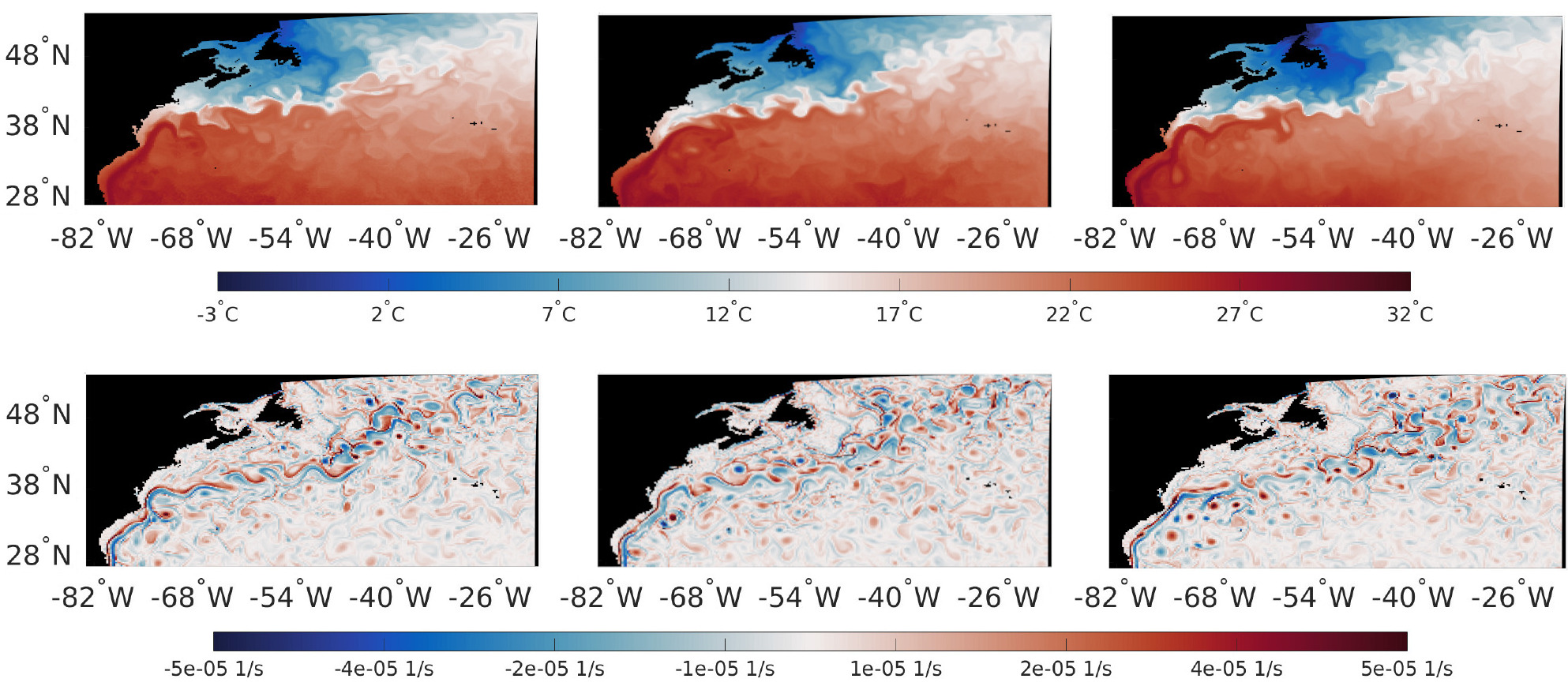}
\caption{Shown are typical SST (top row) and SRV (bottom row) fields sampled from the reference JPD.
}
\label{fig:ref_vs_sampled}
\end{figure}

In order to calibrate the hyper-parameters we take $M=N=5$ and minimize $\mathrm{Di}(A,B)$
with respect to $\eta\in[0,1]$. We have found that $\eta=0.01$ gives the lowest discrepancy.
After setting up the hyper-parameters, we compute a probabilistically-reconstructed phase space by sampling twice as much compared to the reference record length~(figure~\ref{fig:ref_vs_sampled}).

The dimensionality of the probabilistically-reconstructed  space is $n\approx30000$. 
One can directly run the HP method in this high-dimensional phase space, or in a space of reduced
dimensionality if greater speed gains are wanted.
In this work, we reduce the phase space dimension
down to $m=200$ by using the EOF-PC decomposition~\cite{Preisendorfer1988};
note that we decompose the full flow field, including its time mean.
It results in a
two-order of magnitude reduction compared with the original dimension; note that the 200 leading EOFs capture 99\% of the total SST variance and 60\% of the total SRV variance.
We compute the EOF-PC decomposition
of SST and SRV separately; a bivariate SST-SRV decomposition
can be computed if one needs to model these fields together.
As an alternative to the bivariate decomposition, one can decompose these fields separately and combine them in
the HP method, as in~\cite{SB2023_J1}.

The space of PCs (denoted as $\mathbf{x}$) can be regarded
as a reduced version of the probabilistically-reconstructed phase space.
Given this reduced space, one can reconstruct a dynamical system governing
the reduced dynamics of observed reference data, as we did in~\cite{SB2022_J1}.
In this study, we use a different idea: the dynamics of PCs is directly modelled in the reduced
space with the HP method and then the HP solution is projected back into
the full-dimensional probabilistically-reconstructed phase space.

In the reduced phase space, equation~\eqref{eq:evolution_eq_nudging} reads as

\begin{equation}
\frac{d\mathbf{z}}{dt}=\frac{1}{M}\sum\limits_{j\in\mathcal{U}_J}\mathbf{F}(\mathbf{x}_j)+
\eta\left(\frac{1}{N}\sum\limits_{i\in\mathcal{U}_I}\mathbf{x}_i-\mathbf{z}(t)\right),\quad
\mathbf{z}(t_0)=\mathbf{z}_0,
\label{eq:pc}
\end{equation}
\noindent
where $\mathcal{U}_I$ and $\mathcal{U}_J$ are the sets of timesteps indexing the discrete reference solution $\mathbf{x}$ in
the neighbourhood of $\mathbf{z}(t)$.
The only difference of~\eqref{eq:pc} compared with equation~\eqref{eq:evolution_eq_nudging}  is that
$\mathbf{x}$ does not represent the reference solution (it would be SST or SRV in this case) but the PCs.
%
%
Having solved equation~\eqref{eq:pc}, we approximate the HP solution, $\mathbf{y}(t)$,
in the probabilistically-reconstructed phase space by using the leading EOF-PC pairs as follows:

\begin{equation}
\mathbf{y}(t)\approx\sum\limits^m_{i=1}z_i(t)\mathbf{E}_i\,,
\label{eq:y}
\end{equation}
with $\mathbf{E}_i$ and $z_i$ being the i-th EOF and PC (computed with the HP method in the reduced space), respectively. Thus, solution $\mathbf{y}(t)$ is the SST (or SRV) computed in the full-dimensional probabilistically-reconstructed phase space.

\begin{figure}[h!]
\hspace*{-2cm}
\includegraphics[scale=0.3]{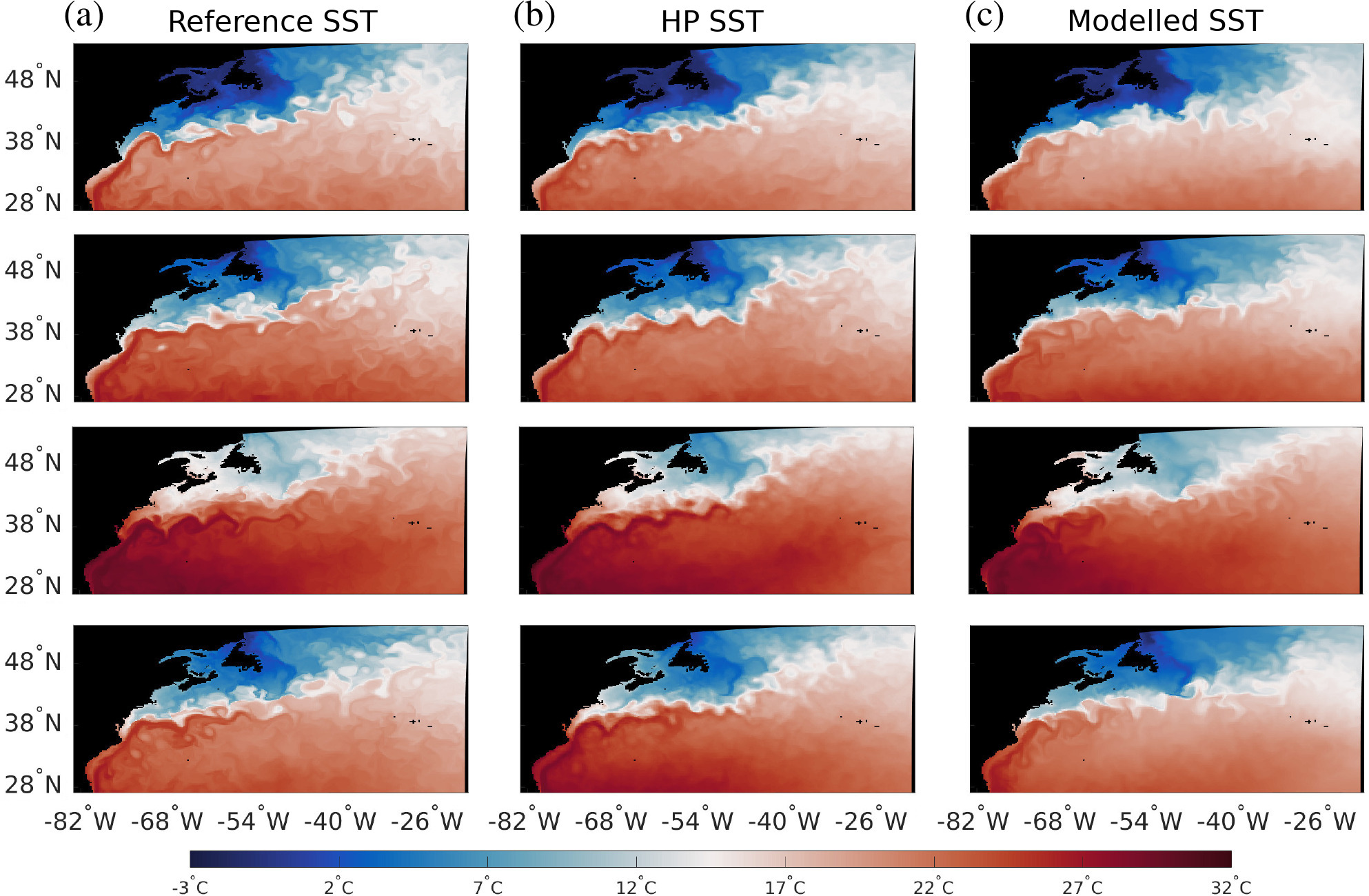}
\caption{Shown are typical SST snapshots at 1/4$^\circ$ resolution for
(a) the reference solution, (b) HP solution, and
(c) modelled solution. The HP solution significantly better represents the Gulf Stream compared with the modelled solution.
}
\label{fig:sst}
\end{figure}

\begin{figure}[h!]
\hspace*{-2cm}
\includegraphics[scale=0.22]{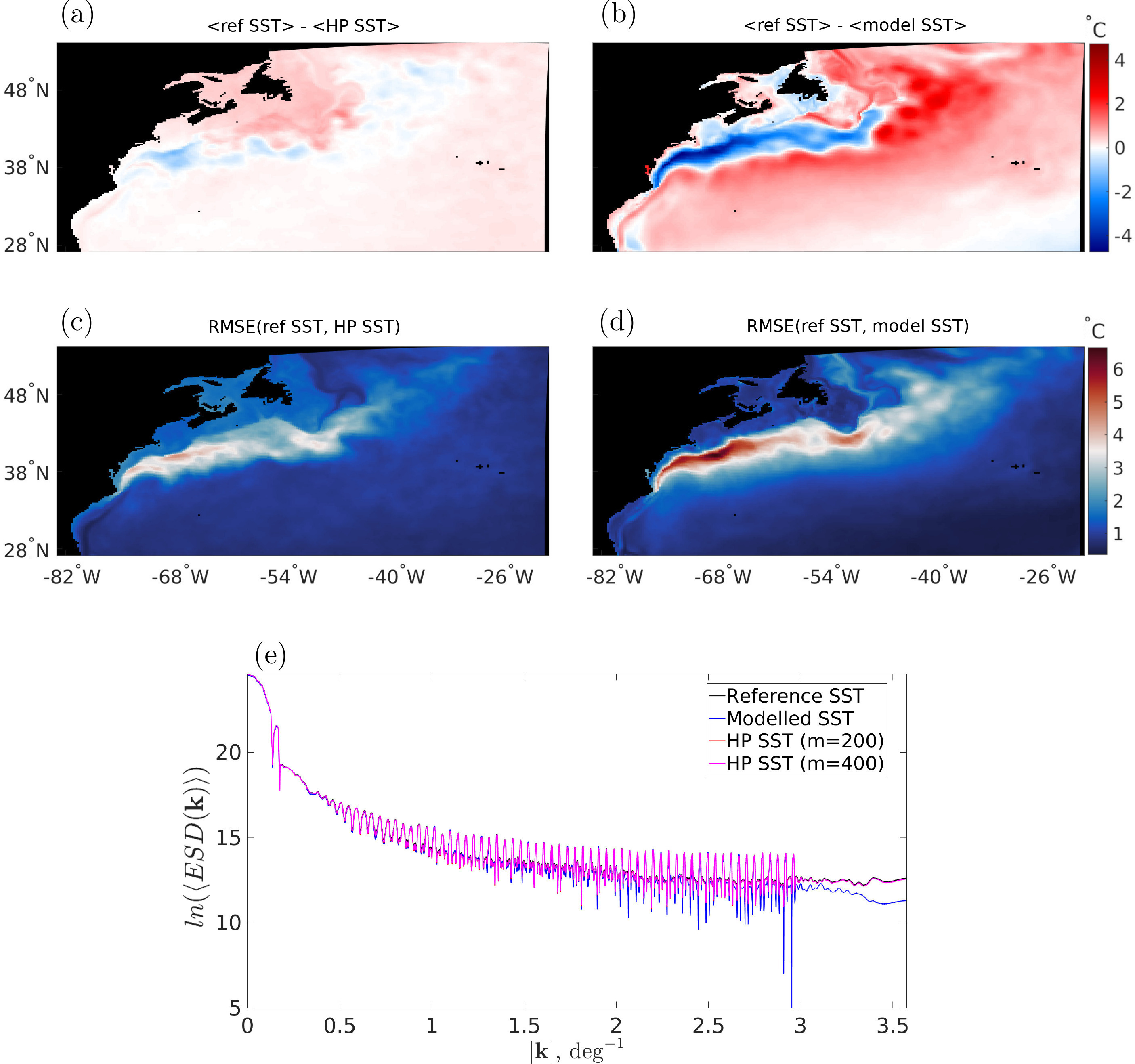}
\caption{Shown are (a) difference between the time mean reference SST, $\langle{\rm ref\, SST}\rangle$,
and HP SST, $\langle{\rm HP\, SST}\rangle$,,
(b) difference between the time mean reference SST and modelled SST, $\langle{\rm model\, SST}\rangle$,
(c) root mean square error, RMSE, between the reference SST and HP SST,
(d) root mean square error, RMSE, between the reference and modelled SST,
(e) the natural logarithm of the time mean energy spectral density $\langle ESD\rangle$ vs the magnitude of the
wave vector $\mathbf{k}$; the units in subplots (a)-(d) and (e) are $[^{\circ}{\rm C}]$
and $[^{\circ}C\cdot {\rm deg}^2]^2$, respectively.
We use $m=200$ in the EOF-PC~decomposition for (a)-(d).
The $\langle ESD\rangle$ of the HP solution (red and magenta)
is closer to the reference (black) than that of the modelled solution (blue).
Note that $\langle ESD\rangle$ for $m=200$ (red) and $m=400$ (blue) are almost the same, as adding more
EOFs does not contribute much to the accuracy of the $\langle ESD\rangle$ of SST.
For illustrative purposes, we apply a locally estimated scatterplot smoothing (also known
as a Savitzky--Golay filter) to smooth the ESD~\cite{StatModels1992}.
}
\label{fig:re_sst}
\end{figure}

\begin{figure}
\hspace*{-2cm}
\includegraphics[scale=0.22]{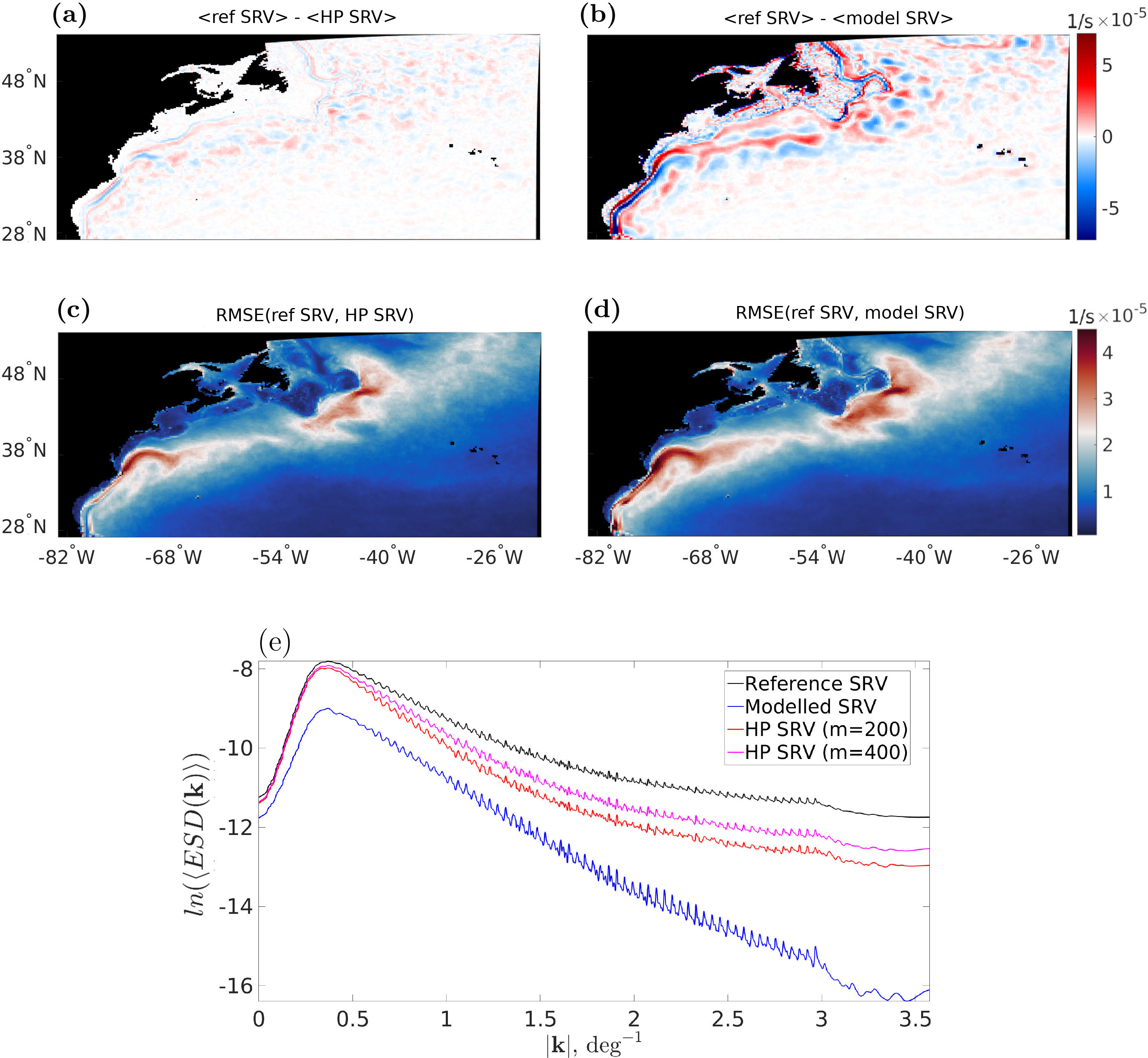}
\caption{The same as in figure~\ref{fig:re_sst} but for SRV;
the units in subplots (a)-(d) and (e) are $[{\rm s}^{-1}]$
and $[{\rm s}^{-1}{\rm deg}^2]^2$, respectively.
Note that adding more EOFs makes the time mean energy spectral density, $\langle ESD\rangle$,
closer to the reference one; compare the $\langle ESD\rangle$ of the HP solution for $m=200$ (red) and $m=400$ (magenta) with the reference (black). The $\langle ESD\rangle$ of the modelled solution (blue) is much
lower compared with that of the HP solution, especially for smaller scales.
}
\label{fig:re_srv}
\end{figure}

We report the results in figure~\ref{fig:sst}, which clearly shows that the HP solution
(figure~\ref{fig:sst}b) much better represents the Gulf Stream than the modelled solution
(figure~\ref{fig:sst}c). These results are also confirmed in figure~\ref{fig:re_sst}. Namely,
the difference between the time mean of the reference solution and HP solution is much smaller
(figure~\ref{fig:re_sst}a)
than that of the modelled solution (figure~\ref{fig:re_sst}b). The root mean square error for the HP solution
(figure~\ref{fig:re_sst}c) is also much lower then the one of the modelled solution (figure~\ref{fig:re_sst}d).
The time mean energy spectral density of the HP solution is closer to the reference than that of
the modelled solution over the whole range of wave numbers (figure~\ref{fig:re_sst}e, compare
the blue and magenta graphs with the black one), and this difference is even more pronounced
for SRV (figure~\ref{fig:re_srv}e).

\begin{figure}
\hspace*{-1.5cm}
\includegraphics[scale=0.3]{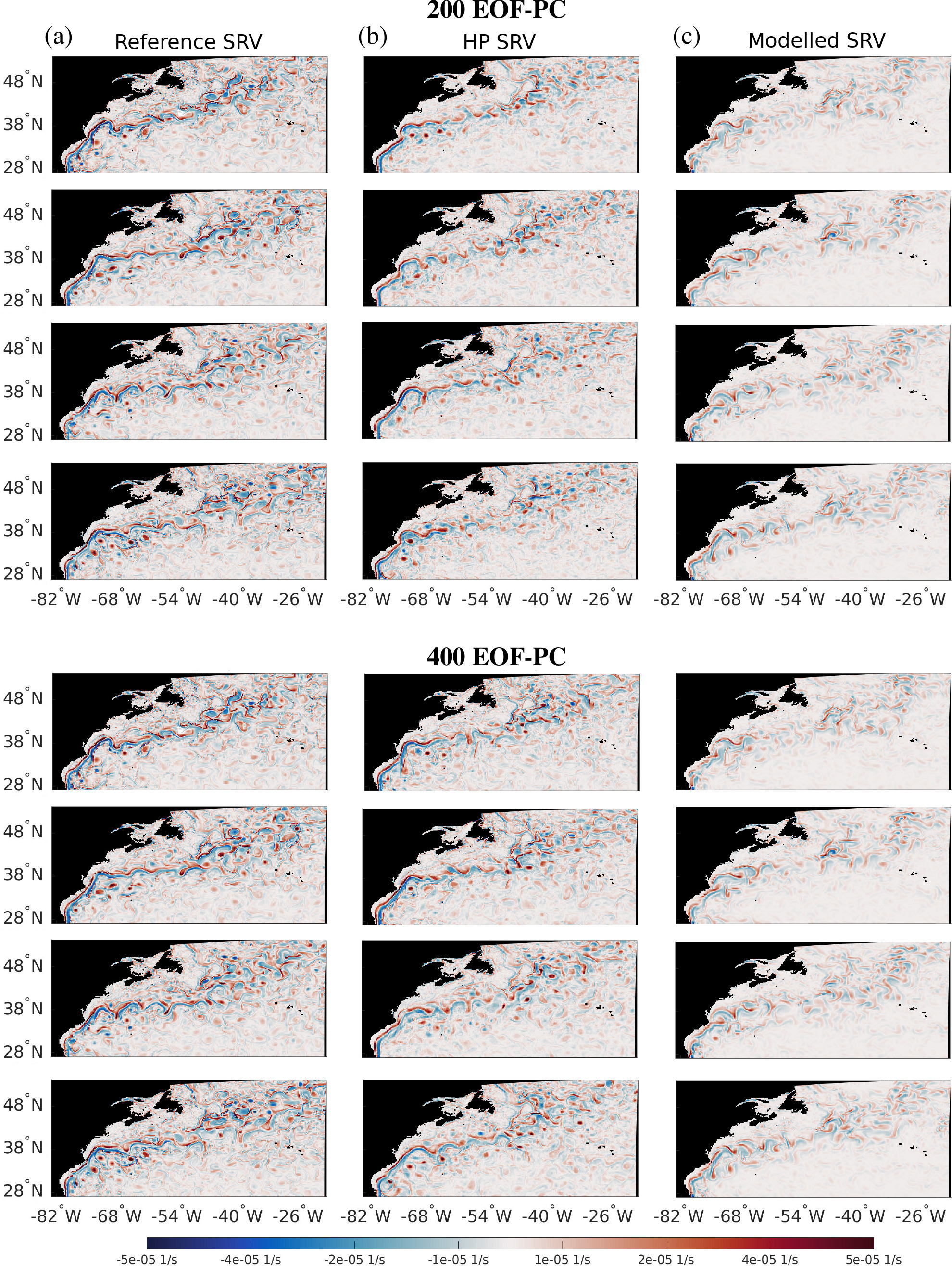}
\caption{The same as in figure~\ref{fig:sst} but for SRV ($m=200$ and $m=400$).
}
\label{fig:srv}
\end{figure}

\ansA{Note that the results shown in figures~\ref{fig:sst}-\ref{fig:re_srv} were obtained using the probabilistically-reconstructed phase space generated by sampling from the estimated JPD. Specifically, the reference dataset (1979-1993) was used to estimate the JPD, from which additional states were sampled to densify the phase space prior to EOF reduction and HP integration.
}

{\bf Remark.}
{\it
The HP solution 'much better'' or ''significantly better'' represents the Gulf Stream
because it reconstructs sharper, more coherent, and spatially accurate SST than
the native 1/4$^\circ$-model. This improvement arises not from enhanced physics but from the
HP method's ability to statistically emulate high-resolution dynamics within
a low-dimensional phase space. Unlike the modelled solution (which suffers from resolution-induced
diffusion, under-resolved mesoscale processes, and smeared fronts) the HP solution inherits
and preserves fine-scale structure resolved at 1/4$^\circ$. In this sense,
'much better'' or ''significantly better'' means that the HP trajectory remains geometrically and statistically
consistent with the reference solution that the 1/4$^\circ$-model cannot represent.
}


\newpage
More insights about how the HP solution differs from the modelled solution can be seen
in the SRV fields (figure~\ref{fig:srv}). The Gulf Stream computed with the HP method (figure~\ref{fig:srv}b)
is much stronger than the one computed with the 1/4$^\circ$ model (figure~\ref{fig:srv}c). Moreover, the HP
solution is teemed with vortices (like the reference one), while the vortex dynamics is significantly
inhibited in the modelled solution as it is also reflected by the error plots in figure~\ref{fig:re_srv},
as well as by the time mean energy spectral density $\langle ESD\rangle$,
\ansA{
and computed from the two-dimensional Fourier transformation of the fields on the longitude-latitude grid as follows.

Let $q(\lambda,\phi,t)$ denote either SST or SRV at time $t$, where longitude $\lambda\in[\lambda_{\min},\lambda_{\max}]$ and latitude $\phi\in[\phi_{\min},\phi_{\max}]$ are measured in degrees.
We denote by $(k_1,k_2)\in\mathbb{Z}^2$ the \emph{integer Fourier indices} associated with the zonal (longitude) and meridional (latitude) directions, respectively.
We define the corresponding wave-vector components (in units of deg$^{-1}$) as
$k_\lambda=k_1/L_\lambda$ and $k_\phi=k_2/L_\phi$, where
$L_\lambda=\lambda_{\max}-\lambda_{\min}$, $L_\phi=\phi_{\max}-\phi_{\min}$.
%
%
The wave-vector magnitude is
$|\mathbf{k}|=\sqrt{k_\lambda^2+k_\phi^2}$.
For a given magnitude $|\mathbf{k}|$, we group Fourier modes into a ring defined by
$
\mathrm{ring}(|\mathbf{k}|)
=
\left\{(k_1,k_2)\in\mathbb{Z}^2:\
\sqrt{k_\lambda^2+k_\phi^2}=|\mathbf{k}|
\right\}
$,
i.e. $\mathrm{ring}(|\mathbf{k}|)$ contains all discrete Fourier modes having the same magnitude $|\mathbf{k}|$. 

Let $\widehat{q}(k_1,k_2,t)$ denote the 2D discrete Fourier transform coefficient of $q(\lambda,\phi,t)$ at mode $(k_1,k_2)$. 
For each snapshot $t$, we define the energy spectral density as

\[
\mathrm{ESD}(\mathbf{k},t):=\sum_{(k_1,k_2)\in \mathrm{ring}(|\mathbf{k}|)}\left|\widehat{q}(k_1,k_2,t)\right|^2.
\]

\noindent
The time mean energy spectral density is then

\[
\langle \mathrm{ESD}(\mathbf{k})\rangle=\frac{1}{T}\sum_{n=1}^{T}\mathrm{ESD}(\mathbf{k},t_n),
\]
where $T$ is the number of available snapshots $\{t_n\}_{n=1}^{T}$ in the period used for averaging.

Note that the small-scale oscillations observed for $|\mathbf{k}|>1$ in
figures~\ref{fig:re_sst},\ref{fig:re_srv},\ref{fig:srv_spectrum_corrupted_data}
represent artifacts of spectral estimation, and are not interpreted as physically meaningful spectral features.
}

SRV is more sensitive to the number of EOFs compared with SST, i.e. the more EOFs are used, the more accurate the
$\langle ESD\rangle$ is. It is because SRV fields contain much more small-scale features (compared with SST), which therefore require more EOFs to be accurately represented.
Although doubling the number of PCs ($m=400$) results in a more pronounced Gulf Stream dynamics and
larger population of vortices (figure~\ref{fig:srv}), as well as in a more accurate representation
of $\langle ESD\rangle$ (figure~\ref{fig:re_srv}e),
its contribution to the time mean flow and the root mean square error
is small (not shown).

\subsection{Damaged reference data}
\begin{figure}
\hspace*{-1cm}
\includegraphics[scale=0.34]{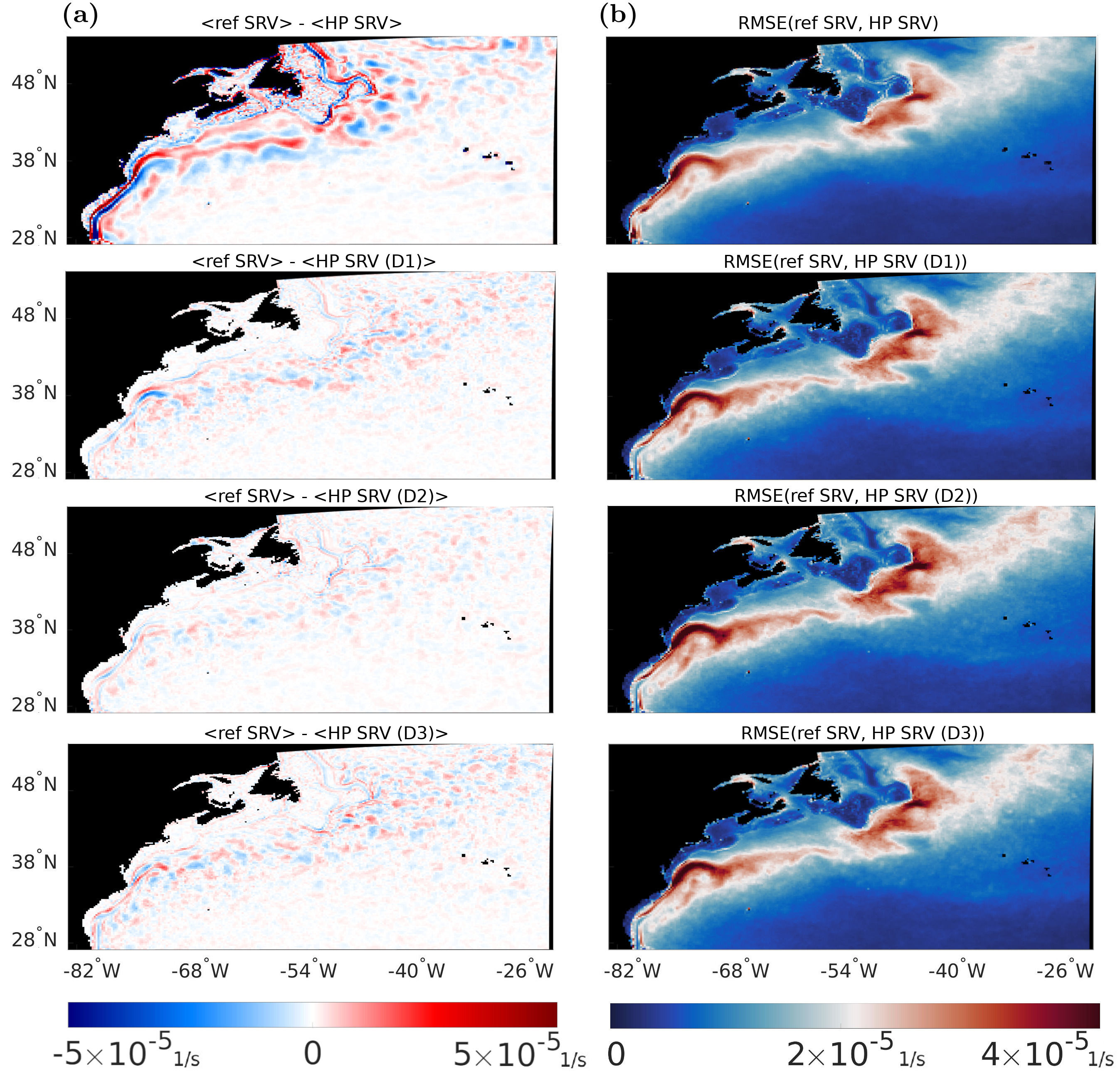}
\caption{Shown are (a) difference between the time mean reference SRV and modelled SRV, as well as
HP SRV (D1), HP SRV (D2), HP SRV (D3) (top to bottom);
(b) root mean square error between the reference SRV and modelled SRV, as well as
HP SRV (D1), HP SRV (D2), HP SRV (D3) (top to bottom).
We use m=200 in the EOF-PC~decomposition for (a) and (b).
The time mean SRV of the HP solution computed from the damaged data ((a) -- second to fourth row) is more
accurate than that of the modelled solution ((a) -- top), while the root mean square error of the HP solution
is on par with that of the modelled solution (b).
}
\label{fig:re_srv_corrupted_data}
\end{figure}

\begin{figure}
\hspace*{-0.25cm}
\includegraphics[scale=0.17]{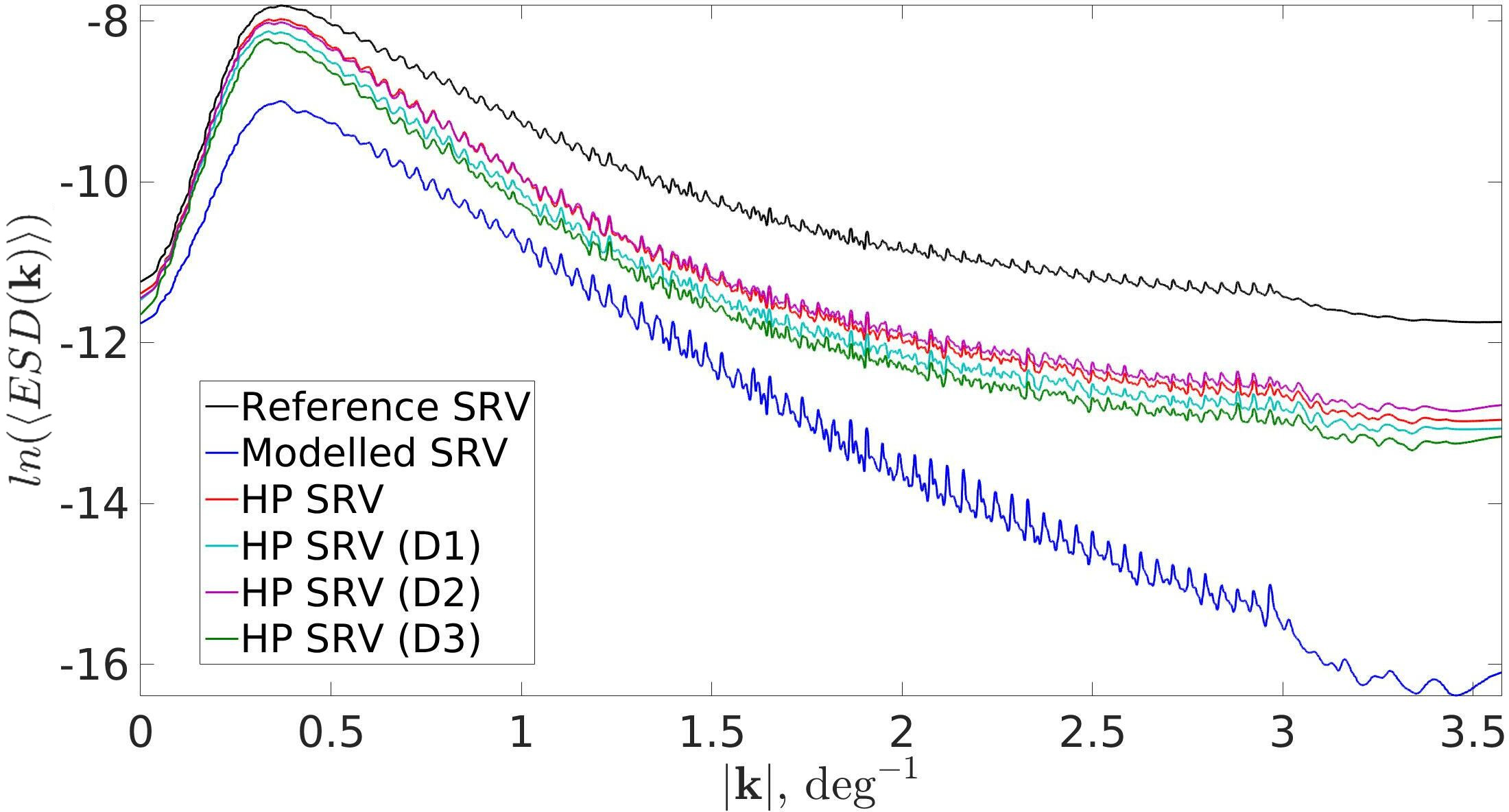}
\caption{
Shown is the natural logarithm of the time mean energy spectral density
$\langle ESD\rangle$ (in units of $[{\rm s}^{-1}\cdot{\rm deg}^2]^2$) vs the magnitude of the
wave vector $\mathbf{k}$.
The $\langle ESD\rangle$ of the HP solution (red) and HP solutions computed from the damaged reference data sets (D1, D2, D3)
are closer to the reference solution (black) than that of the modelled solution (blue).
As in figure~\ref{fig:re_srv_corrupted_data} we use m=200 in the EOF-PC~decomposition.
}
\label{fig:srv_spectrum_corrupted_data}
\end{figure}

Damaged reference data sets are frequently encountered in observations used for data assimilation and reanalysis in comprehensive ocean models.  There is an extensive toolkit of methods currently in circulation to repair damaged data so that it
could be of further use in computations. As we have seen in the example of Chua's circuit,
the HP method in probabilistically-reconstructed phase spaces can work directly with unreliable reference data sets thus
requiring no extra tools to cope with this issue. In this section, we study how damaged reference data from the NEMO model impacts the HP method. Similar to the scenario with Chua's circuit, we examine the reference phase space damaged
by randomly missed states (D1), gaps (D2), and cuts (D3). Each damaged data set is of the size of the validation
period to ensure its length is the same as the one used in runs with undamaged data above.
We only focus on the surface relative vorticity, as it is more sensitive to errors than temperature. The results show that
regardless of the nature of damage, the proposed methodology outperforms the modelled solution by providing a more accurate representation of the time-mean flow
(figure~\ref{fig:re_srv_corrupted_data}), the root mean square error of the HP solution is comparable to that of the modelled solution.
Furthermore, the energy spectrum of the HP solution (across all types of damage) is closer to the reference spectrum compared with the one of the modelled solution (figure~\ref{fig:srv_spectrum_corrupted_data}). These findings allow us to conclude that the proposed methodology can be used as a dynamic interpolation method for repairing damaged observational data.

The HP method, in its standard form, is not designed for within-field spatial
gap-filling, which however can be present in real observations (for example, some
regions can be masked by clouds). It assumes complete phase-space inputs at each time step and operates
on the temporal evolution of full system states, rather than performing grid-level
inpainting in physical space. While the HP method can, in principle, be adapted to
handle partial fields (e.g., by operating on local patches or integrating with a
statistical imputation step), such extensions are beyond the scope of the present
study. Here we focus on demonstrating the HP method for working in phase spaces
with damaged structure.

\clearpage
\section{Conclusions and Discussion}
In this study we have proposed to use the joint probability distribution (JPD) to probabilistically
reconstruct the phase space when the original reference phase space (computed from numerical solutions, observations, or both) is too sparse
for \ansA{the HP approach} to be effective. The probabilistically-reconstructed phase space is calculated by sampling
from the JPD, which provides an additional source of data from the reference distribution.
The method ensures that the reference phase space and the reconstructed space share the same reference distribution.
We have shown how the probabilistic reconstruction works on the example of Chua's system and then successfully applied it
regionally within the context of the global ocean model NEMO.

\ansA{
{\bf On the scope and limitations of HP.} The proposed methodology operates within the phase space defined by the HP solution computed in the probabilistically-reconstructed reference phase space. It is important to distinguish between these two objects. The probabilistically-reconstructed phase space is obtained by sampling from the reference data distribution and may still contain voids, particularly in regions where the original dataset lacked global support (such voids cannot be fully recovered by JPD sampling if entire regions of the attractor were never observed).

In contrast, the phase space traced by the HP solution is dynamically continuous: since the HP trajectory is obtained by integrating equations~\eqref{eq:evolution_eq_nudging} or~\eqref{eq:pc}, it does not consist of isolated sampled points and therefore does not exhibit discrete voids. In this sense, the HP phase space may appear “larger” or more connected than the probabilistically-reconstructed cloud. However, its evolution remains intrinsically tied to the reference phase space, because at every time step
the evolution of the image point relies on neighbourhood statistics computed from the reference dataset. The method therefore does not construct a global parametric model of the dynamics. Instead, it reconstructs trajectories using the reference phase space. Its generalization properties are therefore determined by the coverage of
that phase space.
\begin{itemize}
    \item \textbf{Generalization to different initial conditions.}
    If the initial condition lies within (or sufficiently close to) the reference phase space, the HP method can generate dynamically consistent trajectories. In this sense, the emulator generalizes across initial conditions belonging to the same dynamical regime represented in the reference data.

    \item \textbf{Generalization to different physical parameters.}
    The method does not extrapolate well beyond the parameter regime represented in the reference dataset. If the underlying physical parameters change in such a way that the attractor geometry undergoes significant structural changes, the reference phase space must be recomputed or augmented accordingly.

    \item \textbf{Out-of-distribution states.}
    If the trajectory enters regions far away from the reference phase space, the HP method cannot reliably reconstruct the dynamics, as the neighbourhood information becomes unrepresentative.
\end{itemize}
In this sense, the HP emulator should be viewed as a nonparametric, reference-phase-space-conditioned dynamical reconstruction tool rather than a globally trained parametric surrogate. Its predictive skill depends on phase-space coverage rather than parametric training. \\
}

Despite these limitations, the proposed methodology enables the computation of the HP solution in the probabilistically-reconstructed
(and also reduced) phase space, which proves to be more accurate than the 1/4$^\circ$-NEMO simulation.
However, if the reference distribution changes over time (e.g. the reference solution drifts away from the phase space, used to compute
the reference distribution) then the HP solution cannot follow these changes.

We have also found that the HP method can be used directly in reduced phase spaces,
thus leading to several-orders-of-magnitude acceleration (compared to the 1/4$^\circ$-resolution NEMO run).
This acceleration can be translated into using larger ensembles of HP solutions for probabilistic predictions.
Note that we compare only the degrees of freedom utilized in the HP method
and in the NEMO model to get the solution at $1/4^\circ$ resolution.
It goes without saying that the NEMO model evolves much more variables than the HP method, but in this comparison we pretend
that NEMO needs only $\sim$30000 grid points to compute the solution, while the HP method uses only 200 in the compressed phase space,
and more importantly gives a solution which is closer to the reference one than that of the $1/4^\circ$ NEMO simulation.
%

\ansA{
The computational cost of the proposed methodology is dominated by the offline preprocessing stage (PDF estimation and EOF decomposition). The HP trajectory integration in the reduced space is computationally inexpensive, involving only nearest-neighbour searches and local averaging.
}

We have studied how the proposed method performs on damaged reference data
(for which we considered three cases:randomly-missed states, gaps, and cuts) and found
that the method works well even when sampling from the JPD cannot
patch up voids in the reference phase space.
It allows us to conclude that the HP method in probabilistically-reconstructed phase
spaces can be used as a fast reanalysis method in which the complex dynamics
of a comprehensive ocean model is replaced with the HP solution,
as well as a dynamic interpolation method to fill gaps in observational data.

The primary utility of the proposed methodology is to serve as a fast ocean, atmospheric, or ocean-atmospheric
emulator in the reference phase space, or as an alternative to parameterisations in comprehensive ocean and ocean-atmospheric models.
We believe that it demonstrates the strong potential and it can be of use
in operational ocean and ocean-atmospheric models.

When modelling climate changes with data-driven methods, one might assume
that the reference data set could be a limiting factor. While it is true, it only becomes
a concern if future climate states fall outside the reference phase space.
Since it is a priori unknown whether future climate states are located in the reference phase space,
it may be advisable to use this methodology as an alternative to parameterisations
in comprehensive models for near-term climate predictions (NTCP), instead of
using it solely as a forecasting tool.
As with ensemble forecasts generated by physics-based models,
an ensemble of HP-based emulations, each initialized from slightly different image points
in phase space, will yield distinct trajectories and thus define a probabilistic
description of the flow dynamics. However, the interpretation of this ensemble differs
fundamentally from that of ensembles produced by physics-based models.
In traditional physics-based models, ensemble members divergence arises from nonlinear
instabilities, chaotic sensitivity to initial conditions, model error, variability
in external forcing, etc. -- all governed by dynamical equations. In contrast, HP ensembles reflect
geometric sensitivity to the local structure of the reference phase space. Variability between
HP members originates from how slightly perturbed initial conditions explore different regions of
the phase space. Consequently, the HP ensemble spread reflects
data-induced uncertainty, not dynamical chaos or forcing-driven divergence.\\

Although the HP method is formally time-agnostic and operates entirely within the geometry of a
reference phase space, it can, under certain conditions, act as a forecast system.
Specifically, if future states of the system lie within the statistical and geometric support
of the reference phase space (i.e. they are already geometrically encoded into the reference phase
space) then HP can evolve towards them. In this case, the ensemble reflects statistically
admissible futures.
As long as the evolution remains within the reference phase space,
HP ensembles will closely resemble
physics-based model ensembles in terms of time-evolving statistics such as mean, spread,
spatial and temporal patterns. \\

However, while HP ensembles are computationally
inexpensive and statistically consistent with sampled dynamics, they offer a geometric interpolation
of uncertainty, rather than a physics-driven future forecast.
HP cannot extrapolate beyond that; it lacks any mechanism to explore
regions of phase space not already explicitly or implicitly represented in the reference data.
%
This limitation is not unique to HP. It applies broadly to data-driven emulators that lack
explicit knowledge of governing equations or external forcings. Such methods can produce
internally consistent trajectories and ensemble spreads within the statistical manifold
defined by the training data, but cannot properly generalize beyond it or respond to changes
in physical boundary conditions or forcings that inject the trajectory out of the reference
phase space.
More research is needed to assess prediction skills and achievable forecast range of the
method alone and in combination with high-resolution forecasting systems.

\section*{Declaration of competing interest}
The authors declare that they have no known competing financial interests or personal relationships that could have appeared to influence the work reported in this paper.

\section*{Data availability}
\noindent
The output of the global 1/4$^\circ$ (ORCA025-N4001) and 1/12$^\circ$ (ORCA0083-N006) resolution NEMO model
used in this study is available on JASMIN ( https://jasmin.ac.uk/users/access/ ) from the following locations:
/gws/nopw/j04/nemo\_vol1/ORCA025-N401  and\\
\noindent
/gws/nopw/j04/nemo\_vol1/ORCA0083-N006~.
Simulation data sets produced and analysed in this study
as well as the software used in the analyses can be found in~\cite{Shevchenko2025_zenodo2,Shevchenko2025_zenodo1}.

\section*{Acknowledgments}
The author thanks the Natural Environment Research Council for the support of this work through
the project ATLANTIS (P11742), as well as Andrew Coward and Chris Wilson for the production of
and help with the NEMO datasets, respectively.
The author thanks unknown referees for valuable comments and suggestions, which
helped improve the manuscript

%



\bibliographystyle{elsarticle-num}
\bibliography{refs}



%
%
%

\end{document}